\documentclass[journal,draftclsnofoot,onecolumn,12pt]{IEEEtran}

\usepackage{cite}
\usepackage{amsthm,amssymb,amsmath,graphicx,multirow,color,amsfonts}%
\usepackage[update,prepend]{epstopdf}

\usepackage[latin1]{inputenc}
\usepackage{tikz}
\usepackage{bbm} 
\usepackage{pdfpages}
\usepackage{multirow}
\usepackage{subfig}
\usepackage{comment}
\allowdisplaybreaks 

\usepackage{setspace}	
 \usepackage{graphicx}

\usepackage{algorithm,algorithmic}
\usepackage{multicol}

\usepackage[justification=centering]{caption}
\usepackage{textcomp}
\usepackage{psfrag}
\usepackage{arydshln}
\usepackage{url}
\usepackage{soul}
\usepackage{graphicx,color}
\usepackage[nolist]{acronym}
\usepackage{algorithm,algorithmic} 
\usepackage{mathtools,lipsum}
\usepackage{cuted}
\usepackage{flushend}

\usepackage{amsmath}

\usepackage[capitalise]{cleveref}
\Crefname{equation}{Eq.\!}{Eqs.\!}
\Crefname{figure}{Fig.\!}{Figs.\!}
\Crefname{tabular}{Tab.\!}{Tabs.\!}
\Crefname{section}{Section\!}{Sections.\!}

\def\nb0{{\mathbf{0}}}
\def\nb1{{\mathbf{1}}}

\newtheorem{lemma}{Lemma}

\newtheorem{definition}{Definition}

\newtheorem{theorem}{Theorem}

\newtheorem{corollary}{Corollary}

\newtheorem{remark}{Remark}

\newenvironment{sequation}{
\begin{equation}\small}{\end{equation}
}

\begin{document}
\graphicspath{{./Figures/}}
	\begin{acronym}

\acro{5G-NR}{5G New Radio}
\acro{3GPP}{3rd Generation Partnership Project}
\acro{ABS}{aerial base station}
\acro{AC}{address coding}
\acro{ACF}{autocorrelation function}
\acro{ACR}{autocorrelation receiver}
\acro{ADC}{analog-to-digital converter}
\acrodef{aic}[AIC]{Analog-to-Information Converter}     
\acro{AIC}[AIC]{Akaike information criterion}
\acro{aric}[ARIC]{asymmetric restricted isometry constant}
\acro{arip}[ARIP]{asymmetric restricted isometry property}

\acro{ARQ}{Automatic Repeat Request}
\acro{AUB}{asymptotic union bound}
\acrodef{awgn}[AWGN]{Additive White Gaussian Noise}     
\acro{AWGN}{additive white Gaussian noise}

\acro{APSK}[PSK]{asymmetric PSK} 

\acro{waric}[AWRICs]{asymmetric weak restricted isometry constants}
\acro{warip}[AWRIP]{asymmetric weak restricted isometry property}
\acro{BCH}{Bose, Chaudhuri, and Hocquenghem}        
\acro{BCHC}[BCHSC]{BCH based source coding}
\acro{BEP}{bit error probability}
\acro{BFC}{block fading channel}
\acro{BG}[BG]{Bernoulli-Gaussian}
\acro{BGG}{Bernoulli-Generalized Gaussian}
\acro{BPAM}{binary pulse amplitude modulation}
\acro{BPDN}{Basis Pursuit Denoising}
\acro{BPPM}{binary pulse position modulation}
\acro{BPSK}{Binary Phase Shift Keying}
\acro{BPZF}{bandpass zonal filter}
\acro{BSC}{binary symmetric channels}              
\acro{BU}[BU]{Bernoulli-uniform}
\acro{BER}{bit error rate}
\acro{BS}{base station}
\acro{BW}{BandWidth}
\acro{BLLL}{ binary log-linear learning }

\acro{CP}{Cyclic Prefix}
\acrodef{cdf}[CDF]{cumulative distribution function}   
\acro{CDF}{Cumulative Distribution Function}
\acrodef{c.d.f.}[CDF]{cumulative distribution function}
\acro{CCDF}{complementary cumulative distribution function}
\acrodef{ccdf}[CCDF]{complementary CDF}               
\acrodef{c.c.d.f.}[CCDF]{complementary cumulative distribution function}
\acro{CD}{cooperative diversity}

\acro{CDMA}{Code Division Multiple Access}
\acro{ch.f.}{characteristic function}
\acro{CIR}{channel impulse response}
\acro{cosamp}[CoSaMP]{compressive sampling matching pursuit}
\acro{CR}{cognitive radio}
\acro{cs}[CS]{compressed sensing}                   
\acrodef{cscapital}[CS]{Compressed sensing} 
\acrodef{CS}[CS]{compressed sensing}
\acro{CSI}{channel state information}
\acro{CCSDS}{consultative committee for space data systems}
\acro{CC}{convolutional coding}
\acro{Covid19}[COVID-19]{Coronavirus disease}

\acro{DAA}{detect and avoid}
\acro{DAB}{digital audio broadcasting}
\acro{DCT}{discrete cosine transform}
\acro{dft}[DFT]{discrete Fourier transform}
\acro{DR}{distortion-rate}
\acro{DS}{direct sequence}
\acro{DS-SS}{direct-sequence spread-spectrum}
\acro{DTR}{differential transmitted-reference}
\acro{DVB-H}{digital video broadcasting\,--\,handheld}
\acro{DVB-T}{digital video broadcasting\,--\,terrestrial}
\acro{DL}{DownLink}
\acro{DSSS}{Direct Sequence Spread Spectrum}
\acro{DFT-s-OFDM}{Discrete Fourier Transform-spread-Orthogonal Frequency Division Multiplexing}
\acro{DAS}{Distributed Antenna System}
\acro{DNA}{DeoxyriboNucleic Acid}

\acro{EC}{European Commission}
\acro{EED}[EED]{exact eigenvalues distribution}
\acro{EIRP}{Equivalent Isotropically Radiated Power}
\acro{ELP}{equivalent low-pass}
\acro{eMBB}{Enhanced Mobile Broadband}
\acro{EMF}{ElectroMagnetic Field}
\acro{EU}{European union}
\acro{EI}{Exposure Index}
\acro{eICIC}{enhanced Inter-Cell Interference Coordination}

\acro{FC}[FC]{fusion center}
\acro{FCC}{Federal Communications Commission}
\acro{FEC}{forward error correction}
\acro{FFT}{fast Fourier transform}
\acro{FH}{frequency-hopping}
\acro{FH-SS}{frequency-hopping spread-spectrum}
\acrodef{FS}{Frame synchronization}
\acro{FSsmall}[FS]{frame synchronization}  
\acro{FDMA}{Frequency Division Multiple Access}

\acro{GA}{Gaussian approximation}
\acro{GF}{Galois field }
\acro{GG}{Generalized-Gaussian}
\acro{GIC}[GIC]{generalized information criterion}
\acro{GLRT}{generalized likelihood ratio test}
\acro{GPS}{Global Positioning System}
\acro{GMSK}{Gaussian Minimum Shift Keying}
\acro{GSMA}{Global System for Mobile communications Association}
\acro{GS}{ground station}
\acro{GMG}{ Grid-connected MicroGeneration}

\acro{HAP}{high altitude platform}
\acro{HetNet}{Heterogeneous network}

\acro{IDR}{information distortion-rate}
\acro{IFFT}{inverse fast Fourier transform}
\acro{iht}[IHT]{iterative hard thresholding}
\acro{i.i.d.}{independent, identically distributed}
\acro{IoT}{Internet of Things}                      
\acro{IR}{impulse radio}
\acro{lric}[LRIC]{lower restricted isometry constant}
\acro{lrict}[LRICt]{lower restricted isometry constant threshold}
\acro{ISI}{intersymbol interference}
\acro{ITU}{International Telecommunication Union}
\acro{ICNIRP}{International Commission on Non-Ionizing Radiation Protection}
\acro{IEEE}{Institute of Electrical and Electronics Engineers}
\acro{ICES}{IEEE international committee on electromagnetic safety}
\acro{IEC}{International Electrotechnical Commission}
\acro{IARC}{International Agency on Research on Cancer}
\acro{IS-95}{Interim Standard 95}

\acro{KPI}{Key Performance Indicator}

\acro{LEO}{low earth orbit}
\acro{LF}{likelihood function}
\acro{LLF}{log-likelihood function}
\acro{LLR}{log-likelihood ratio}
\acro{LLRT}{log-likelihood ratio test}
\acro{LoS}{Line-of-Sight}
\acro{LRT}{likelihood ratio test}
\acro{wlric}[LWRIC]{lower weak restricted isometry constant}
\acro{wlrict}[LWRICt]{LWRIC threshold}
\acro{LPWAN}{Low Power Wide Area Network}
\acro{LoRaWAN}{Low power long Range Wide Area Network}
\acro{NLoS}{Non-Line-of-Sight}
\acro{LiFi}[Li-Fi]{light-fidelity}
 \acro{LED}{light emitting diode}
 \acro{LABS}{LoS transmission with each ABS}
 \acro{NLABS}{NLoS transmission with each ABS}

\acro{MB}{multiband}
\acro{MC}{macro cell}
\acro{MDS}{mixed distributed source}
\acro{MF}{matched filter}
\acro{m.g.f.}{moment generating function}
\acro{MI}{mutual information}
\acro{MIMO}{Multiple-Input Multiple-Output}
\acro{MISO}{multiple-input single-output}
\acrodef{maxs}[MJSO]{maximum joint support cardinality}                       
\acro{ML}[ML]{maximum likelihood}
\acro{MMSE}{minimum mean-square error}
\acro{MMV}{multiple measurement vectors}
\acrodef{MOS}{model order selection}
\acro{M-PSK}[${M}$-PSK]{$M$-ary phase shift keying}                       
\acro{M-APSK}[${M}$-PSK]{$M$-ary asymmetric PSK} 
\acro{MP}{ multi-period}
\acro{MINLP}{mixed integer non-linear programming}

\acro{M-QAM}[$M$-QAM]{$M$-ary quadrature amplitude modulation}
\acro{MRC}{maximal ratio combiner}                  
\acro{maxs}[MSO]{maximum sparsity order}                                      
\acro{M2M}{Machine-to-Machine}                                                
\acro{MUI}{multi-user interference}
\acro{mMTC}{massive Machine Type Communications}      
\acro{mm-Wave}{millimeter-wave}
\acro{MP}{mobile phone}
\acro{MPE}{maximum permissible exposure}
\acro{MAC}{media access control}
\acro{NB}{narrowband}
\acro{NBI}{narrowband interference}
\acro{NLA}{nonlinear sparse approximation}
\acro{NLOS}{Non-Line of Sight}
\acro{NTIA}{National Telecommunications and Information Administration}
\acro{NTP}{National Toxicology Program}
\acro{NHS}{National Health Service}

\acro{LOS}{Line of Sight}

\acro{OC}{optimum combining}                             
\acro{OC}{optimum combining}
\acro{ODE}{operational distortion-energy}
\acro{ODR}{operational distortion-rate}
\acro{OFDM}{Orthogonal Frequency-Division Multiplexing}
\acro{omp}[OMP]{orthogonal matching pursuit}
\acro{OSMP}[OSMP]{orthogonal subspace matching pursuit}
\acro{OQAM}{offset quadrature amplitude modulation}
\acro{OQPSK}{offset QPSK}
\acro{OFDMA}{Orthogonal Frequency-division Multiple Access}
\acro{OPEX}{Operating Expenditures}
\acro{OQPSK/PM}{OQPSK with phase modulation}

\acro{PAM}{pulse amplitude modulation}
\acro{PAR}{peak-to-average ratio}
\acrodef{pdf}[PDF]{probability density function}                      
\acro{PDF}{probability density function}
\acrodef{p.d.f.}[PDF]{probability distribution function}
\acro{PDP}{power dispersion profile}
\acro{PMF}{probability mass function}                             
\acrodef{p.m.f.}[PMF]{probability mass function}
\acro{PN}{pseudo-noise}
\acro{PPM}{pulse position modulation}
\acro{PRake}{Partial Rake}
\acro{PSD}{power spectral density}
\acro{PSEP}{pairwise synchronization error probability}
\acro{PSK}{phase shift keying}
\acro{PD}{power density}
\acro{8-PSK}[$8$-PSK]{$8$-phase shift keying}
\acro{PPP}{Poisson point process}
\acro{PCP}{Poisson cluster process}
 
\acro{FSK}{Frequency Shift Keying}

\acro{QAM}{Quadrature Amplitude Modulation}
\acro{QPSK}{Quadrature Phase Shift Keying}
\acro{OQPSK/PM}{OQPSK with phase modulator }

\acro{RD}[RD]{raw data}
\acro{RDL}{"random data limit"}
\acro{ric}[RIC]{restricted isometry constant}
\acro{rict}[RICt]{restricted isometry constant threshold}
\acro{rip}[RIP]{restricted isometry property}
\acro{ROC}{receiver operating characteristic}
\acro{rq}[RQ]{Raleigh quotient}
\acro{RS}[RS]{Reed-Solomon}
\acro{RSC}[RSSC]{RS based source coding}
\acro{r.v.}{random variable}                               
\acro{R.V.}{random vector}
\acro{RMS}{root mean square}
\acro{RFR}{radiofrequency radiation}
\acro{RIS}{Reconfigurable Intelligent Surface}
\acro{RNA}{RiboNucleic Acid}
\acro{RRM}{Radio Resource Management}
\acro{RUE}{reference user equipments}
\acro{RAT}{radio access technology}
\acro{RB}{resource block}

\acro{SA}[SA-Music]{subspace-augmented MUSIC with OSMP}
\acro{SC}{small cell}
\acro{SCBSES}[SCBSES]{Source Compression Based Syndrome Encoding Scheme}
\acro{SCM}{sample covariance matrix}
\acro{SEP}{symbol error probability}
\acro{SG}[SG]{sparse-land Gaussian model}
\acro{SIMO}{single-input multiple-output}
\acro{SINR}{signal-to-interference plus noise ratio}
\acro{SIR}{signal-to-interference ratio}
\acro{SISO}{Single-Input Single-Output}
\acro{SMV}{single measurement vector}
\acro{SNR}[\textrm{SNR}]{signal-to-noise ratio} 
\acro{sp}[SP]{subspace pursuit}
\acro{SS}{spread spectrum}
\acro{SW}{sync word}
\acro{SAR}{specific absorption rate}
\acro{SSB}{synchronization signal block}
\acro{SR}{shrink and realign}

\acro{tUAV}{tethered Unmanned Aerial Vehicle}
\acro{TBS}{terrestrial base station}

\acro{uUAV}{untethered Unmanned Aerial Vehicle}
\acro{PDF}{probability density functions}

\acro{PL}{path-loss}

\acro{TH}{time-hopping}
\acro{ToA}{time-of-arrival}
\acro{TR}{transmitted-reference}
\acro{TW}{Tracy-Widom}
\acro{TWDT}{TW Distribution Tail}
\acro{TCM}{trellis coded modulation}
\acro{TDD}{Time-Division Duplexing}
\acro{TDMA}{Time Division Multiple Access}
\acro{Tx}{average transmit}

\acro{UAV}{Unmanned Aerial Vehicle}
\acro{uric}[URIC]{upper restricted isometry constant}
\acro{urict}[URICt]{upper restricted isometry constant threshold}
\acro{UWB}{ultrawide band}
\acro{UWBcap}[UWB]{Ultrawide band}   
\acro{URLLC}{Ultra Reliable Low Latency Communications}
         
\acro{wuric}[UWRIC]{upper weak restricted isometry constant}
\acro{wurict}[UWRICt]{UWRIC threshold}                
\acro{UE}{User Equipment}
\acro{UL}{UpLink}

\acro{WiM}[WiM]{weigh-in-motion}
\acro{WLAN}{wireless local area network}
\acro{wm}[WM]{Wishart matrix}                               
\acroplural{wm}[WM]{Wishart matrices}
\acro{WMAN}{wireless metropolitan area network}
\acro{WPAN}{wireless personal area network}
\acro{wric}[WRIC]{weak restricted isometry constant}
\acro{wrict}[WRICt]{weak restricted isometry constant thresholds}
\acro{wrip}[WRIP]{weak restricted isometry property}
\acro{WSN}{wireless sensor network}                        
\acro{WSS}{Wide-Sense Stationary}
\acro{WHO}{World Health Organization}
\acro{Wi-Fi}{Wireless Fidelity}

\acro{sss}[SpaSoSEnc]{sparse source syndrome encoding}

\acro{VLC}{Visible Light Communication}
\acro{VPN}{Virtual Private Network} 
\acro{RF}{Radio Frequency}
\acro{FSO}{Free Space Optics}
\acro{IoST}{Internet of Space Things}

\acro{GSM}{Global System for Mobile Communications}
\acro{2G}{Second-generation cellular network}
\acro{3G}{Third-generation cellular network}
\acro{4G}{Fourth-generation cellular network}
\acro{5G}{Fifth-generation cellular network}	
\acro{gNB}{next-generation Node-B Base Station}
\acro{NR}{New Radio}
\acro{UMTS}{Universal Mobile Telecommunications Service}
\acro{LTE}{Long Term Evolution}

\acro{QoS}{Quality of Service}
\end{acronym}
	
\newcommand{\SAR} {\mathrm{SAR}}
\newcommand{\WBSAR} {\mathrm{SAR}_{\mathsf{WB}}}
\newcommand{\gSAR} {\mathrm{SAR}_{10\si{\gram}}}
\newcommand{\Sab} {S_{\mathsf{ab}}}
\newcommand{\Eavg} {E_{\mathsf{avg}}}
\newcommand{\ft}{f_{\textsf{th}}}
\newcommand{\alphatf}{\alpha_{24}}

\title{
Resident Population Density-Inspired Deployment of K-tier Aerial Cellular Network
}

\author{
Ruibo Wang, Mustafa A. Kishk, {\em Member, IEEE} and Mohamed-Slim Alouini, {\em Fellow, IEEE}
\thanks{Ruibo Wang and Mohamed-Slim Alouini are with King Abdullah University of Science and Technology (KAUST), CEMSE division, Thuwal 23955-6900, Saudi Arabia. Mustafa A. Kishk is with the Department of Electronic Engineering, National University of Ireland, Maynooth, W23 F2H6, Ireland. (e-mail: ruibo.wang@kaust.edu.sa; mustafa.kishk@mu.ie;
slim.alouini@kaust.edu.sa). 
}
\vspace{-4mm}
}

\maketitle

\begin{abstract}
Using \acp{UAV} to enhance network coverage has proven a variety of benefits compared to terrestrial counterparts. One of the commonly used mathematical tools to model the locations of the UAVs is stochastic geometry (SG). However, in the existing studies, both users and UAVs are often modeled as homogeneous point processes. In this paper, we consider an inhomogeneous \ac{PPP}-based model for the locations of the users that captures the degradation in the density of active users as we move away from the town center. In addition, we propose the deployment of aerial vehicles following the same inhomogeneity of the users to maximize the performance. In addition, a multi-tier network model is also considered to make better use of the rich space resources. Then, the analytical expressions of the coverage probability for a typical user and the total coverage probability are derived. Finally, we optimize the coverage probability with limitations of the total number of UAVs and the minimum local coverage probability. Finally we give the optimal UAV distribution parameters when the maximum overall coverage probability is reached.
\end{abstract}


\section{Introduction}
In the next generation mobile network (5G, beyond 5G), UAVs have many application scenarios \cite{sekander2018multi,li2018uav,9504595}, among which UAV-aided ubiquitous coverage becomes an important topic \cite{zeng2016wireless}. Because UAVs are easy to deploy, highly mobile, and have 3D deployment, they are often used to build temporary or dynamic networks and provide ubiquitous coverage. Especially, UAV is widely used to relieve the pressure of large crowds gathering in small areas \cite{zeng2016wireless,alzenad2018fso}. They are proven to provide reliable system coverage in hot spots and provide additional system performance \cite{kishk2020aerial}. Due to the demand for high rate signals, multi-tier vertical heterogeneous networks (VHetNet) is proposed to make use of space resources in city centers \cite{huo2019distributed,wang2019optimized}. 
\par
One of the main unanswered questions in the realm of UAV-enabled wireless networks is where and how high the UAVs should be deployed \cite{UAV_SG}. The common assumption in SG-based literature is that the user's spatial distribution is homogeneous. Consequently, existing literature typically assumes that the density of the UAVs is spatially invariant. However, according to recent studies on resident population densities, a more proper assumption would be for the density of the users (and consequently the UAVs) drops as their distance from the town center increases \cite{herculea2016straight}. Analyzing the influence of such a setup on the wireless network's performance is the main objective of this paper. More details on the contributions of this paper are provided later in Sec.~\ref{contribution}.

\subsection{Related Work}
SG is a powerful mathematical method of analyzing communication networks with irregular topology \cite{haenggi2012stochastic}. Furthermore, the SG framework is suitable for modeling and analyzing devices in motion, such as UAVs, cars \cite{steinmetz2015stochastic}, and LEO satellites \cite{wang2022stochastic,wang2022conditional}. The SG-based analytical results of the network coverage probability can provide accurate approximations to the actual network \cite{satellite_Niloofar, wang2022evaluating}. Next, the authors in \cite{al2014optimal} proposed an air-to-ground line-of-sight (LoS) probability model suitable for town centers. In this model, the probability of the UAV being blocked by the building decreases with the increase of the elevation angle of the UAV to the typical user. This model divides UAVs into LoS UAVs and non-line-of-sight (NLoS) UAVs. Since the model is related to density, area, and height of building \cite{al2014modeling}, it is suitable for various scenarios. Based on the LoS probability model, there has been some literature on UAV networking in town centers \cite{wang2018modeling,kishk2020aerial}. 

\par
In the existing research, some resident population density models have been considered. Different user distributions in several urban environments are proposed in \cite{galkin2019stochastic}. A disjoint clustered model for large resident population density is set up in \cite{bao2018user}. A central model is provided in \cite{herculea2016straight} and is adopted in this paper. In the central model, user density decreases with the distance from the user to the center. However, the above articles pay more attention to the modeling of users, while the UAVs are simply deployed. UAVs are deployed as a homogeneous PPP in \cite{herculea2016straight,bao2018user}, while the locations of UAVs are determined by clustering in \cite{galkin2019stochastic}. Therefore, the deployment of UAVs is also worth exploring. However, with regard to analyzing downlink network coverage performance, changing the distribution of UAVs brings much more difficulty in technical derivation than changing the distribution of users. Given that UAVs form a homogeneous PPP, the downlink coverage performance of users at any location is the same. Nevertheless, when the density of the UAV is not constant, the distributions of the distance between the serving UAV and the interfering UAV to the user are different for the users at different locations, which makes the analysis challenging. 

\par
To effectively utilize the deployable space of UAVs, developing the vertical deployment mode of UAVs is also worth studying, in addition to designing the horizontal distribution of UAVs. Based on the SG framework, authors in \cite{wang2019optimized,UAV_SG,kim2019multi,bao2015handoff,talgat2020nearest,talgat2020stochastic} have put forward multi-tier VHetNets consisting of ground \acp{BS}, UAVs and \acp{HAP} and \ac{LEO}-satellites. The above researches all introduced the concept of association probability to describe the probability of users choosing a communication device in a specific tier (instead of other tiers) to provide services. Unfortunately, the above analytical framework is unsuitable for our study because the UAVs are not uniformly distributed in our paper. Designing a different method to obtain the association probability is another challenge.

\subsection{Contribution}\label{contribution}
The contributions of this paper can be summarized as follows:
\begin{itemize}
    \item We study a resident population density-inspired model of the urban area. The density of users decreases with the distance to the town center. UAVs follow a similar distribution to the distribution of users and are deployed at different altitudes with different densities. 
    \item We derive the analytical result of coverage probability under the specific model and prove that it is consistent with the Monte-Carlo simulation. In addition, the existing coverage probability analysis framework is extended to data rate and energy efficiency.
    \item The coverage performance of multi-tier networks and single-tier networks are compared. We also compare the coverage performance of the population density-inspired distribution and the homogeneous distribution of UAV.
    \item By adjusting the distribution of UAVs in each tier, we optimize the coverage probability under different user distributions. Furthermore, remarks on the parameter design criterion for UAV distribution are given.
\end{itemize}

\section{System Model}
\subsection{Network Model}

\begin{figure*}[h]
	\centering
	\includegraphics[width=0.8\linewidth]{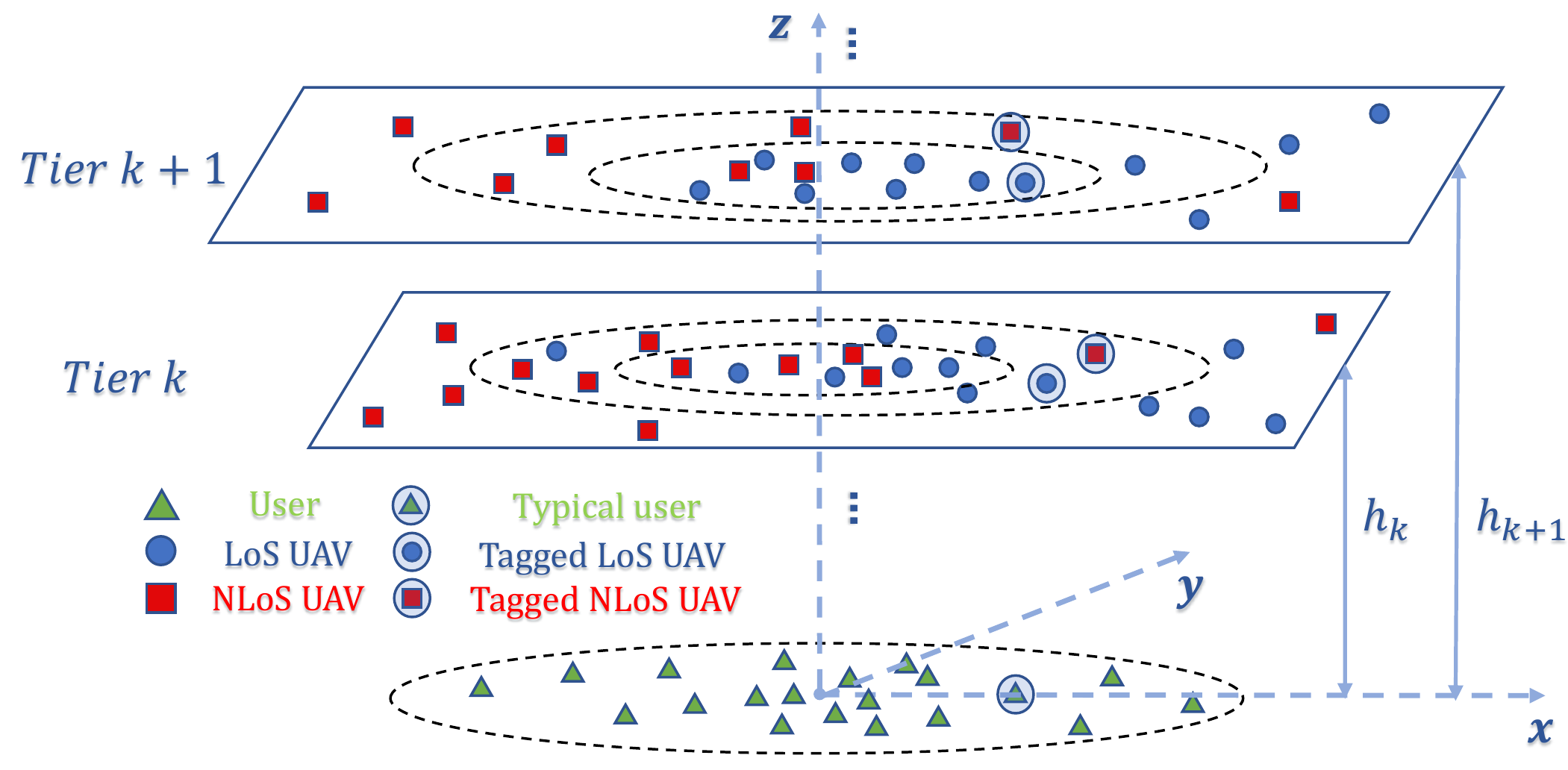}
	\caption{Illustration of the system model.}
	\label{fig:Illustration of the system model}
\end{figure*}

As shown in Fig.~\ref{fig:Illustration of the system model}, we consider a scenario in which ground users are distributed according to an inhomogeneous PPP, which is inspired by the resident population distribution model proposed in \cite{herculea2016straight}. We assume that the UAVs in the VHetNet are deployed based on the ground users density and, hence, their locations also follow an inhomogeneous PPP. Assuming that the center of the town is located at the origin, the densities of the users and the \acp{UAV} near the origin are relatively high, while the density goes down as we move away from the origin. Assume the users are located at the ground, with horizontal distance $z_u$ to the origin, the density distribution of the users ${\Lambda_u}\left(z_u\right)$ can be represented as follows,
\begin{equation}
\label{Lambda_u}
    {\Lambda_u}\left(z_u\right) = {\lambda _u}{e^{ - {\beta _u}z_u}},
\end{equation}
where $\lambda_u$ determines the total density of the plane, $\beta_u$ is a measure of homogeneity. When the value of $\beta_u$ is large, the users are spatially condensed at the origin. When $\beta_u = 0$ the process degenerates to a homogeneous \ac{PPP}. Without loss of generality, we focus on a typical user located on the positive $X$-axis.

We assume that $K$ tiers of \acp{UAV} are distributed at a set of some fixed heights $h_k$ independently. Their location distribution of each tier form a 2D inhomogeneous PPP, denoted by ${\Phi _k}\mathop  = \limits^\Delta  \left\{ {{x _{i,k}}} \right\}$ where ${{x _{i,k}}}$ refers to the 3D location of UAV $i$ in tier $k$. We prefer polar coordinates $\left( z_i, \theta _i, h_k \right)$ to represent ${x _{i,k}}$, where $z_i$ is the horizontal distance between the \ac{UAV} and the origin, $\theta _i$ is the angle between the $X$-axis and the line which connects the projection of the \ac{UAV} and the origin. Because designers tend to place more UAVs in densely populated areas, it is reasonable to assume that they will be in the same distribution as the users. Thus, in tier $k$, the density distribution ${\Lambda_{\rm{UAV},k}}$ can be described as,
\begin{equation}
\label{Lambda_U}
   {\Lambda _{\rm{UAV},k}}\left(z_i\right) = {\lambda _{k}}{e^{ - {\beta _{k}}z_i}},
\end{equation}
where $\lambda_k$ and $\beta_k$ are parameters of the \acp{UAV} in tier k, which have the same meaning as the users' parameters in the density distribution. Furthermore, we assume that each tier of \acp{UAV} have the same transmitting power ${\rho_k}$. The quad ${T_{\rm{k}}} = \left\{ {{h_k},{\lambda _k},{\beta _k},{\rho_k}} \right\},k = 1,2,...,K$ is used to represent the parameters of the ${k^{th}}$ tier.

\subsection{Channel Model}


To model the air-to-ground channel between a user and a UAV, we need to take into consideration the LoS and NLoS scenarios  \cite{al2014optimal}. Considering a \ac{UAV} in tier $k$, given the horizontal distance $z$ between the UAV's projection on the ground and the user, the probability of setting up an LoS link between the typical user and the UAV is \cite{al2014optimal}, \cite{alzenad20173},
\begin{equation}
\label{PLoSk}
    P_k^{{\rm{LoS}}}\left( {\rm{z}} \right) = \frac{1}{{1 + a\exp \left( { - b\left( {\frac{{180}}{\pi }{{\tan }^{ - 1}}\left( {\frac{{{h_k}}}{z}} \right){-a}} \right)} \right)}},
\end{equation}
where $a$ and $b$ are environment-dependent parameters. From the perspective of the typical user, the inhomogeous PPP process corresponding to the $K$-tier UAVs can be split into two disjoint PPPs, that is ${\Phi _k}{=}{\Phi _{{\rm{LoS}},k}} \cup {\Phi _{{\rm{NLoS}},k}}$ and ${\Phi _{{\rm{LoS}},k}} \cap {\Phi _{{\rm{NLoS}},k}} = \phi $, where ${\Phi _{{\rm{LoS}},k}}$ and ${\Phi _{{\rm{NLoS}},k}}$ denote the set of UAVs which establish LoS and NLoS conditions for the typical user respectively, $\phi$ is the empty set.

In this article, UAVs with an LoS link to the typical user are abbreviated as LoS UAVs, while the rest are abbreviated as NLoS UAVs. Therefore, the 3D VHetNet is split into $2K$ disjoint two-dimensional PPPs, with the density $P_k^{Q}\, {\Lambda_{\rm{UAV},k}}$, where $Q=\left\{ {{\rm{LoS}},{\rm{NLoS}}} \right\}$. After LoS and NLoS states are defined, the channel fading model can be established, which is described by small-scale fading and large-scale fading.

For small scale fading, we denote channel fading power gains in terms of independent random variables $G_{\rm{LoS}}$ and $G_{\rm{NLoS}}$, under LoS and NLoS conditions for the typical user, respectively. In order to represent several fading scenarios, Nakagami-m fading is experienced with shape parameters and scale parameters $(m_{\rm{LoS}}, \frac{1}{m_{\rm{LoS}}})$ and $(m_{\rm{NLoS}}, \frac{1}{m_{\rm{NLoS}}})$ for LoS and NLoS links, respectively. As a result, the \ac{PDF} of the power gains $G_Q$ is given by \cite{nakagami}
\begin{equation}
    {f_{G_Q}}\left(g\right) = \frac{{{m_Q}^{{m_Q}}{r^{{m_Q} - 1}}}}{{\Gamma \left( {{m_Q}} \right)}}{e^{ - {m_Q} \, g}},
\end{equation}
where $\Gamma \left( {{m_Q}} \right) = \int_0^\infty  {{x^{{m_Q} - 1}}{e^{ - x}}dx}$ is the Gamma function, $Q=\left\{\rm{LoS},\rm{NLoS}\right\}$. 

For large scale fading, $\eta_{\rm{LoS}}$ and $\eta_{\rm{NLoS}}$ are mean additional gain for LoS and NLoS transmissions \cite{al2014optimal}, with ${\eta _{\rm{LoS}}} > {\eta _{\rm{NLoS}}}$ satisfied. Combining small scale and large scale fading, the received power of the typical user, transmitted by a \ac{UAV} in tier k, is given by,
\begin{align}
\label{received power}
{S_k}\left(r\right) =
\left\{
 	\begin{array}{ll}
{ {\eta_{\rm{LoS}}}{\rho_k}{G _{\rm{LoS}}}{r^{ - {\alpha _{\rm{LoS}}}}}}&{{\rm{in\,case\,of\,LoS}}}\\
{\eta_{\rm{NLoS}}}{\rho_k}{G _{\rm{NLoS}}}{r^{ - {\alpha _{\rm{NLoS}}}}}&{{\rm{in\,case\,of\,NLoS}}}
	\end{array}.
	\right.
\end{align}
where ${\rho_k}$ is the transmission power of \acp{UAV} in tier $k$, 
$\alpha_{\rm{LoS}}$ and $\alpha_{\rm{NLoS}}$ are path-loss exponents for LoS and NLoS transmissions, with ${\alpha _{\rm{LoS}}} <  {\alpha _{\rm{NLoS}}}$ satisfied, $r$ is the Euclidean distance between the \ac{UAV} and the typical user, which is computed by polar coordinates $\left( z_i , \theta _i, h_k \right)$ of the UAV and the horizontal distance $z_u$ from the user to the origin
\begin{equation}
\label{r:user to UAV}
    r_{{x_{i},k}} = \sqrt {{{({z_i}{\rm{cos}}{\theta _i} - {z_u})}^2} + {{({z_i}{\rm{sin}}{\theta _i})}^2} + h_k^2}.
\end{equation}

\subsection{Interference}\label{Association-Interference}
In each tier, the closest LoS and NLoS UAVs are called tagged UAVs. According to the strongest average received power association strategy \cite{UAV_SG}, the typical user will associate with the UAV with the strongest average received power among these $2K$ tagged UAVs. Furthermore, denote the location of the associated UAV as $x_o$. Note that the typical user may not associate with the closest UAV, since in different tiers, the transmitted power of the UAVs is different.


In urban areas where UAVs are densely distributed, it is necessary to consider interference between UAVs.  Considering a worst-case scenario, except for the associated UAV, other LoS and NLoS UAVs interfere with the typical user, which we refer to as interfering UAVs. Given that the horizontal distance from the typical user to the origin is $z_u$ and the associated UAV is in tier $j$, the total interference can be expressed as a function of $r_{x_o}$, which is the Euclidean distance between the associated UAV and the typical user,
\begin{equation}
\label{Interference}
\begin{split}
    {I_j}\left( {r_{x_o},{z_u} } \right) & =  \sum\limits_{k = 1}^K \Big( \sum\limits_{{x_{i,k}} \in {\Phi _{{\rm{LoS}},k}}\backslash \left\{ {{x_o}} \right\}} {{\eta _{{\rm{LoS}}}}{\rho _k}{G_{{\rm{LoS}}}}r_{{x_{i},k}}^{ - {\alpha _{{\rm{LoS}}}}}}  + \sum\limits_{{x_{i,k}} \in {\Phi _{{\rm{NLoS}},k}}\backslash \left\{ {{x_o}} \right\}} {{\eta _{{\rm{NLoS}}}}{\rho _k}{G_{{\rm{NLoS}}}}r_{{x_{i},k}}^{ - {\alpha _{{\rm{NLoS}}}}}}  \Big).
\end{split}
\end{equation}

As shown in the above formulation, the value of the total interference is related to $z_u$, $r_{x_o}$ and the tier where the associated UAV is located. $r_{x_o}$ and the transmission power of tier $j$ determine the received power from the associated UAV.

\subsection{Performance Analysis}

Assuming the associated UAV is located in tier $j$, the instantaneous \ac{SINR} at the typical user is given by the following equation,
\begin{align}
{\rm{SINR}} =
\left\{
 	\begin{array}{ll}
{\frac{{{\eta _{{\rm{LoS}}}}{\rho_j}{G_{{\rm{LoS}}}}r_{{x_{o}}}^{^{ - {\alpha _{{\rm{LoS}}}}}}}}{{{I_j}\left({{r_{x_o}}\left| {{z_u}} \right.} \right) + {\sigma ^2}}}}&{{x_o} \in {\Phi_{{\rm{LoS}},k}}}\\
{\frac{{{\eta _{{\rm{NLoS}}}}{\rho_j}{G_{{\rm{NLoS}}}}r_{{x_o}}^{^{ - {\alpha _{{\rm{NLoS}}}}}}}}{{{I_j}\left( {{r_{{x_o}}}\left| {{z_u}} \right.} \right) + {\sigma^2}}}}&{{x_o} \in {\Phi _{{\rm{NLoS}},k}}}
	\end{array}.
	\right.
\end{align}
where $\sigma^2$ is the \ac{AWGN} power, and ${I_j}\left( {r_{x_o}\left| {{z_u}} \right.} \right)$ is the total interference power. 

The reliability of the service provided can be evaluated by the average performance of \ac{SINR}. In consequence, the coverage probability, which represents the probability that the system can provide reliable connections is defined as the probability that the SINR is greater than a predefined threshold $\gamma$: 
\begin{equation}\label{Definition of coverage probability}
{P^C} \mathop= \limits^\Delta  \mathbb{P}\left[ {\rm{SINR} > \gamma } \right].
\end{equation}

\section{Problem Formulation}
In this section, our objective is to obtain the analytical expression of coverage probability. From the definition of coverage probability in (\ref{Definition of coverage probability}), we know that distinguishing the associated UAV and the interfering UAVs is a prerequisite for computing the coverage probability. Taking an LoS associated UAV for example, the following steps are used to obtain the analytic expression for the coverage probability: (\romannumeral1) derive the \ac{PDF} of the distance distribution of tagged UAV in each tier $k$, denoted as ${f_{{\rm{RLoS}},k}}\left( {r,z_u} \right)$, (\romannumeral2) calculate the probability of the tagged UAV in tier $k$ being associated with the typical user, defined as association probability  $P_{{\rm{LoS}},k}^A\left( {r,z_u} \right)$, with ${f_{{\rm{RLoS}},k}}\left( {r,z_u} \right)P_{{\rm{LoS}},k}^A\left( {r,z_u} \right)$ being the PDF of the distance between the associated UAV in tier $k$ and the typical user, (\romannumeral3) for a specific distance $r$ between the typical user and its associated UAV, denote the probability that the \ac{SINR} is greater than the threshold $\gamma$ as the conditional coverage probability ${P^C}\left( {\gamma,{z_u}\left| r \right.} \right)$. The average coverage probability is obtained by taking the expectation of the conditional coverage probability ${P^C}\left( {\gamma,r,{z_u}} \right)$ with respect to the PDF of $r$ in step (\romannumeral2). Steps (\romannumeral1) - (\romannumeral3) will be explained in Sec.~ \ref{Distance Distribution of Tagged UAV} Sec.~\ref{Association Probabilities}, and Sec.~\ref{Coverage Probability}, respectively.

\subsection{Nearest Interfering UAVs}
A clear understanding of the location range of interfering UAVs is necessary when analyzing tagged or associated drone distribution. Obviously, for tier $k$, the nearest interfering UAV should locate at a distance larger than $h_k$ for the typical user. Another possible lower bound of the distance for an interfering UAV is to ensure a lower average receiving power than the associated UAV. In the following lemmas, the distance between the typical user and the nearest interfering UAVs is given.

\begin{lemma}\label{nearest interfering LoS}
Given that the typical user is associated with a LoS UAV located at distance $r$ in tier $j$, the closest interfering LoS and NLoS UAV in tier $k$ are at least at distances ${d_{{\rm{LoS}}{\rm{ - }}{\rm{LoS}},j,k}}\left(r\right)$ and ${d_{{\rm{LoS}}{\rm{ - }}\rm{NLoS},j,k}}\left(r\right)$ , given by
\begin{equation}
\label{LoS-LoS,j,k}
    {d_{{\rm{LoS - LoS}},j,k}}\left( r \right) = \max \left\{ {{h_k},{{\left( {\frac{{{\rho_k}}}{{{\rho_j}}}} \right)}^{\frac{1}{{{\alpha _{{\rm{LoS}}}}}}}}r} \right\},
\end{equation}
\begin{equation}
\begin{split}
\label{dLoS-NLoS,j,k}
    &d_{{\rm{LoS - NLoS}},j,k}\left( r \right) = \max \left\{ {{h_k},{{\left( {\frac{{{\eta _{{\rm{NLoS}}}}{\rho_k}\mathbb{E}[{G_{{\rm{NLoS}}}}]}}{{{\eta _{{\rm{LoS}}}}{\rho_j}\mathbb{E}[{G_{{\rm{LoS}}}}]}}} \right)}^{\frac{1}{\alpha _{{\rm{NLoS}}}}}}{r^{\frac{\alpha _{{\rm{LoS}}}}{\alpha_{\rm{NLoS}}}}}} \right\}.
\end{split}
\end{equation}
\begin{proof} 
According to the received power in (\ref{received power}), the average received power of the associated UAV at distance$r$in tier $j$ is $S_j = {\eta_{\rm{LoS}}}{\rho_j}{G_{{\rm{LoS}}}}{r^{ - {\alpha _{{\rm{LoS}}}}}}$. The closest interfering LoS UAV in tier $k$ is at least at distances $d_{{\rm{LoS - LoS}},j,k}$, which can be obtained by solving the equality ${\eta_{\rm{LoS}}}{\rho_j}{G_{{\rm{LoS}}}}{r^{ - {\alpha _{{\rm{LoS}}}}}} = {\eta _{{\rm{LoS}}}}{\rho_k}{G_{{\rm{LoS}}}}d_{{\rm{LoS - LoS}},j,k}^{ - {\alpha _{{\rm{LoS}}}}}$. Similarly, for NLoS interfering UAVs, $d_{{\rm{LoS - NLoS}},j,k}$ can be obtained by solving the equality ${\eta_{\rm{LoS}}}{\rho_j}{G_{{\rm{LoS}}}}{r^{ - {\alpha _{{\rm{LoS}}}}}} = {\eta _{{\rm{NLoS}}}}{\rho_k}{G_{{\rm{NLoS}}}}d_{{\rm{LoS - NLoS}},j,k}^{ - {\alpha _{{\rm{NLoS}}}}}$.

\end{proof}
\end{lemma}


\begin{lemma}\label{nearest interfering NLoS}
Given that the typical user is associated with a NLoS UAV located at distance $r$ in tier $j$, the closest interfering LoS and NLoS UAV in tier $k$ are at least at distances ${d_{{\rm{NLoS}}{\rm{ - }}{\rm{LoS}},j,k}}\left(r\right)$ and ${d_{{\rm{NLoS}}{\rm{ - }}\rm{NLoS},j,k}}\left(r\right)$ , given by
\begin{equation}
\begin{split}
\label{dNLoS-LoS,j,k}
    &{d_{{\rm{NLoS - LoS}},j,k}}\left( r \right) = \max \left\{ {{h_k},{{\left( {\frac{{{\eta _{{\rm{LoS}}}}{\rho_k}\mathbb{E}[{G_{{\rm{LoS}}}}]}}{{{\eta _{{\rm{NLoS}}}}{\rho_j}\mathbb{E}[{G_{{\rm{NLoS}}}}]}}} \right)}^{\frac{1}{{{\alpha _{{\rm{LoS}}}}}}}}{r^{\frac{{{\alpha _{{\rm{NLoS}}}}}}{{{\alpha _{{\rm{LoS}}}}}}}}} \right\},
\end{split}
\end{equation}
\begin{equation}
\label{dNLoS-NLoS,j,k}
    {d_{{\rm{NLoS - NLoS}},j,k}}\left( r \right) = \max \left\{ {{h_k},{{\left( {\frac{{{\rho_k}}}{{{\rho_j}}}} \right)}^{\frac{1}{{{\alpha _{{\rm{NLoS}}}}}}}}r} \right\}.
\end{equation}

\begin{proof}
The proof is similar to that of lemma~\ref{nearest interfering LoS}.
\end{proof}
\end{lemma}

In the subsequent analysis, the horizontal distance is more practical than the Euclidean distance in this model. The horizontal distances ${z_{Q,j,k}}\left( r \right)$ corresponding to lemma~\ref{nearest interfering LoS} and lemma~\ref{nearest interfering NLoS} are defined as
\begin{equation}
\label{zQjk}
    {z_{Q,j,k}}\left( r \right) = \sqrt {d_{_{Q,j,k}}^2\left( r \right) - h_k^2},
\end{equation}
where $Q = \{ {\rm{LoS}} - {\rm{LoS}},{\rm{LoS}} - {\rm{NLoS}},{\rm{NLoS}} - {\rm{LoS}},{\rm{NLoS}} - {\rm{NLoS}}\} $.

\subsection{Distance Distribution of Tagged UAV}\label{Distance Distribution of Tagged UAV}

Before deriving the \ac{PDF} of the distance of associated UAV, obtaining the distance distributions of tagged UAVs is necessary. The distance distributions are given in the following lemmas.
\begin{lemma}\label{CDF of Tagged LoS UAV} 
Given the distance between the typical user and the origin is $z_u$, the CDF of the distance between the tagged LoS UAV in tier $k$ and the typical user is given by,
\begin{sequation}
\begin{split}
    &{F_{{\rm{RLoS}},k}}\left( {r,z_u} \right) = 1 - \exp \Bigg( - \int_{{z_u} - \sqrt {{r^2} - h_k^2} }^{{z_u} + \sqrt {{r^2} - h_k^2} } \int_{ - \varphi_{\rm{LoS-LoS}} \left( {l,r,{z_u}} \right)}^{\varphi_{\rm{LoS-LoS}} \left( {l,r,{z_u}} \right)} {v_k^{\rm{LoS}}\left( {z_u},l,\theta \right) \mathsf{d}\theta \mathsf{d}l}   \Bigg),
\end{split}
\end{sequation}
where
\begin{equation}
\label{phi}
    \varphi_{Q,j,k} \left( {l,r,{z_u}} \right) = \arccos \left( {\frac{{{l^2} + z_u^2 - z_{Q,j,k}^2\left( r \right)}}{{2 \, l \, {z_u}}}} \right),
\end{equation}
\begin{equation}
\label{vkQ}
    v_k^Q\left( {z_u},l,\theta \right) = \left| l \right| \, {\Lambda _{\rm{UAV},k}}\left( l \right) \, P_k^{Q}\left( d_{\rm{u2U}}\left( {z_u},l,\theta  \right)\right),
\end{equation}
where $Q = \{ {\rm{LoS}} - {\rm{LoS}},{\rm{LoS}} - {\rm{NLoS}},{\rm{NLoS}} - {\rm{LoS}},{\rm{NLoS}} - {\rm{NLoS}}\} $, the horizontal distances ${z_{Q,j,k}}\left( r \right)$ are defined in (\ref{zQjk}), $\Lambda_{\rm{UAV},k} \left( r \right)$ and ${P_k^{{\rm{LoS}}} \left( z \right) }$ are given in (\ref{Lambda_U}) and (\ref{PLoSk}), respectively. The distance between the potential interfering UAV and the typical user ${d_{\rm{u2U}}}\left( {{z_u},l,\theta}\right)$ in (\ref{vkQ}) is given by,
\begin{equation}
\label{du2U}
    {d_{\rm{u2U}}}\left( {{z_u},l,\theta } \right) = \sqrt {{{\left( {{z_u} - l\cos \theta } \right)}^2} + {{\left( {l\sin \theta } \right)}^2}}.
\end{equation}
\begin{proof}
See Appendix~\ref{app:CDF of Tagged LoS UAV}.
\end{proof}
\end{lemma}


\begin{lemma}\label{CDF of Tagged NLoS UAV} 
Given the distance between the typical user and the origin is $z_u$, the CDF of distance between the tagged NLoS UAV in tier $k$ and the typical user is given by
\begin{sequation}
\begin{split}
    &{F_{{\rm{RNLoS}},k}}\left( {r,z_u} \right) = 1 -  \exp \Bigg(  - \int_{{z_u} - \sqrt {{r^2} - h_k^2} }^{{z_u} + \sqrt {{r^2} - h_k^2} } \int_{ - \varphi_{\rm{NLoS-NLoS}} \left( {l,r,{z_u}} \right)}^{\varphi_{\rm{NLoS-NLoS}} \left( {l,r,{z_u}} \right)} {v_k^{\rm{NLoS}}\left( {z_u},l,\theta \right)\mathsf{d}\theta \mathsf{d}l}   \Bigg),
\end{split}
\end{sequation}
where ${v_k^{Q}\left( {{z_u},l,\theta } \right)}$ and $\varphi_{Q,j,k} \left( {l,r,{z_u}} \right)$ are given in (\ref{vkQ}) and (\ref{phi}), respectively.   
\begin{proof}
The proof is similar to that of Lemma~\ref{CDF of Tagged LoS UAV}, therefore omitted here.
\end{proof}
\end{lemma}

\begin{lemma}\label{PDF of Tagged UAV} 
Given the distance between the typical user and the origin is $z_u$, the PDF of distance between the tagged $Q$ UAV in tier $k$ and the typical user is given by,
\begin{equation}
\begin{split}
\label{sequation: PDF of tagged UAV}
   & {f_{{\rm{R}Q},k}}\left( {r,z_u} \right) =  \exp \Bigg(  - \int_{{z_u} - \sqrt {{r^2} - h_k^2} }^{{z_u} + \sqrt {{r^2} - h_k^2} } \int_{ - \varphi_{Q-Q} \left( {l,r,{z_u}} \right)}^{\varphi_{Q-Q} \left( {l,r,{z_u}} \right)} v_k^Q\left( {z_u},l,\theta \right){\mathsf{d}\theta \mathsf{d}l}   \Bigg)\\
    & \times \Bigg(\int_{{z_u} - \sqrt {{r^2} - h_k^2} }^{{z_u} + \sqrt {{r^2} - h_k^2} }  - \frac{4r \,\mathbbm{1} \left( {r > {h_k}} \right) \, v_k^Q\left( {z_u},l,\varphi_{Q-Q}  \right)}{\sqrt {4 \, {l^2} \, z_u^2 - {{\left( {{l^2} + z_u^2 - {r^2} + h_k^2} \right)}^2}} }\mathsf{d}l \\
    & + \int_{ - \varphi_{Q-Q} \left( {{z_u} + \sqrt {{r^2} - h_k^2} ,r,{z_u}} \right)}^{\varphi_{Q-Q} \left( {{z_u} + \sqrt {{r^2} - h_k^2} ,r,{z_u}} \right)} \frac{r\;v_k^Q\left( {z_u},{z_u + \sqrt {{r^2} - h_k^2}  },\theta \right)}{\sqrt {{r^2} - h_k^2} } \mathsf{d}\theta \\
    & +  \int_{ - \varphi_{Q-Q} \left( {{z_u} - \sqrt {{r^2} - h_k^2} ,r,{z_u}} \right)}^{\varphi_{Q-Q} \left( {{z_u} - \sqrt {{r^2} - h_k^2} ,r,{z_u}} \right)}  \frac{r\;v_k^Q\left( {z_u},{z_u - \sqrt {{r^2} - h_k^2}  },\theta \right)}{\sqrt {{r^2} - h_k^2} } \mathsf{d}\theta \Bigg),\\
\end{split}
\end{equation}
where $\mathbbm{1}\left( {r > {h_k}} \right)$ is an indicator function, its value is $1$ when $r > {h_k}$ is satisfied, otherwise $0$, ${v_k^{Q} \left( {{z_u},l,\theta } \right)}$ and $\varphi_{Q,j,k} \left( {l,r,{z_u}} \right)$ are given in (\ref{vkQ}) and (\ref{phi}), respectively. For a LoS tagged UAV, $Q$ in (\ref{sequation: PDF of tagged UAV}) is replaced with LoS, while $Q$ is replaced with NLoS for an NLoS tagged UAV.
\par
\begin{proof}
See Appendix~\ref{app:PDF of Tagged UAV}.
\end{proof}
\end{lemma}

\subsection{Association Probabilities}\label{Association Probabilities}

Association probability is used to describe the probability that a tagged UAV will eventually be selected as the associated UAV. For the tagged LoS UAV in tier $k$, there will be no LoS UAVs providing stronger power in tier $k$ than the tagged UAV, while the NLoS UAVs in tier $k$ may provide stronger average received power, and in other tiers, both LoS and NLoS UAVs may provide stronger power. As a result, the association probabilities are given in the following lemmas.

\begin{lemma}\label{Association of LoS UAV}
Given the distance between the typical user and the origin is $z_u$, for the LoS tagged UAV from tier $j$ at Euclidean distance $r$ from the typical user, the probability that the typical user is associated with this specific UAV is given by
\begin{equation}
\begin{split}
\label{PALoSj}
    P_{{\rm{LoS}},j}^A\left( {r,z_u} \right) & =  \prod\limits_{k = 1,j \ne k}^K \exp \Bigg( -  \int_{{z_u} - {z_{{\rm{LoS - LoS}},j,k}}\left( r \right)}^{{z_u} + {z_{{\rm{LoS - LoS}},j,k}}\left( r \right)} \int_{ - \varphi_{\rm{LoS-LoS}} \left( {l,r,{z_u}} \right)}^{\varphi_{\rm{LoS-LoS}} \left( {l,r,{z_u}} \right)} {v_k^{{\rm{LoS}}}\left( {{z_u},l,\theta } \right)  \mathsf{d}\theta \mathsf{d}l} \Bigg) \\
    & \times \prod\limits_{k = 1}^K  \exp \Bigg(  -  \int_{{z_u} - {z_{{\rm{LoS - NLoS}},j,k}} \left( r \right)}^{{z_u} + {z_{{\rm{LoS - NLoS}},j,k}}\left( r \right)} \int_{ - \varphi_{\rm{LoS-NLoS}} \left( {l,r,{z_u}} \right)}^{\varphi_{\rm{LoS-NLoS}} \left( {l,r,{z_u}} \right)}  {v_k^{{\rm{NLoS}}} \left( {{z_u},l,\theta } \right)  \mathsf{d}\theta \mathsf{d}l} \Bigg) ,
\end{split}
\end{equation}
where ${v_k^{Q}\left( {{z_u},l,\theta } \right)}$ and $\varphi_{Q,j,k} \left( {l,r,{z_u}} \right)$ are given in (\ref{vkQ}) and (\ref{phi}), respectively. 
\begin{proof}
See Appendix~\ref{app:Association of LoS UAV}.
\end{proof}
\end{lemma}

\begin{lemma}\label{Association of NLoS UAV}
Given the distance between the typical user and the origin is $z_u$, for the NLoS tagged UAV from tier $j$ at Euclidean distance $r$ from the typical user, the probability that the typical user is associated with this specific UAV is given by
\begin{equation}
\label{PANLoSj}
\begin{split}
    P_{{\rm{NLoS}},j}^A & \left( {r,z_u} \right)  = \prod\limits_{k = 1}^K \exp \Bigg(  - \int_{{z_u} - {z_{{\rm{NLoS - LoS}},j,k}}\left( r \right)}^{{z_u} + {z_{{\rm{NLoS - LoS}},j,k}}\left( r \right)} \int_{ - \varphi_{\rm{NLoS-LoS}} \left( {l,r,{z_u}} \right)}^{\varphi_{\rm{NLoS-LoS}} \left( {l,r,{z_u}} \right)} {v_k^{{\rm{LoS}}} \left( {z_u},l,\theta \right)\mathsf{d}\theta \mathsf{d}l}   \Bigg) \\
    & \times \prod\limits_{k = 1,j \ne k}^K \exp  \Bigg( - \int_{{z_u} - {z_{{\rm{NLoS - NLoS}},j,k}}\left( r \right)}^{{z_u} + {z_{{\rm{NLoS - NLoS}},j,k}}\left( r \right)} \int_{ - \varphi _{\rm{NLoS-NLoS}} \left( {l,r,{z_u}} \right)}^{\varphi_{\rm{NLoS-NLoS}} \left( {l,r,{z_u}} \right)} {v_k^{{\rm{NLoS}}} \left( {z_u},l,\theta \right) \mathsf{d}\theta\mathsf{d}l}  \Bigg) ,
\end{split}
\end{equation}
where ${v_k^{Q}\left( {{z_u},l,\theta } \right)}$ and $\varphi_{Q,j,k} \left( {l,r,{z_u}} \right)$ are given in (\ref{vkQ}) and (\ref{phi}), respectively. 
\begin{proof}
The proof is similar to that of Lemma~\ref{Association of LoS UAV}, therefore omitted here.
\end{proof}
\end{lemma}

\subsection{Coverage Probability}\label{Coverage Probability}

As an indispensable intermediate result to enable computing coverage probability, the Laplace Transform of interference is given in the following lemma.

\begin{lemma}\label{LT of interference}
Given that the distance between the typical user and the origin is $z_u$, the Laplace transform of the interference power
conditioned on the associated UAV in tier $j$ with Euclidean distance $r$ from the typical user is given by,
\begin{equation}
\begin{split}
    \label{LIS}
    &{\mathcal{L}_{I_{Q_1,j}}}\left( {s,r,z_u} \right) = \prod\limits_{k = 1}^K \Big[ {\mathcal{L}_{I_{Q_1-\rm{LoS},j,k}}}\left( {s,r,z_u} \right) \times {\mathcal{L}_{I_{Q_1-\rm{NLoS},j,k}}}\left( {s,r,z_u} \right) \Big] ,
\end{split}
\end{equation}
where $Q_1$ is replaced with LoS when the typical user is associated with LoS UAV, $Q_1$ is replaced with NLoS when NLoS UAV is associated, and ${\mathcal{L}_{I_{Q_1-Q_2,j,k}}}\left( {s,r,z_u} \right)$, $Q_2 = \{\rm{LoS}, \rm{NLoS}\}$ is given by,
\begin{equation}
\begin{split}
  {\mathcal{L}_{I_{Q_1-Q_2,j,k}}}  \left( {s,r,z_u} \right)  =  \exp & \Bigg(  - \int_0^{\max \left\{ {0,{z_u} - {z_{Q_1-Q_2,j,k}}\left( r \right)} \right\}} \int_{ - \pi }^\pi  {v_k^{Q_2}\left( {{z_u},l,\theta } \right)  {w_{Q_2,k}}\left( {s,{z_u},l,\theta}\right) \mathsf{d}\theta \mathsf{d}l}   \Bigg) \\
  & \times \exp \Bigg( -  \int_{{z_u} + {z_{Q_1-Q_2,j,k}}\left( r \right)}^{ + \infty } \int_{ - \pi }^\pi  {v_k^{Q_2} \left( {{z_u},l,\theta } \right) {w_{Q_2,k}}\left( {s,{z_u},l,\theta}\right)\mathsf{d}\theta \mathsf{d}l}   \Bigg)\\
  \times \exp \Bigg( - 2 & \int_{{z_u} - {z_{Q_1-Q_2,j,k}}\left( r \right)}^{{z_u} + {z_{Q_1-Q_2,j,k}}\left( r \right)} \int_{\varphi_{Q_1-Q_2,j,k} \left( {l,r,{z_u}} \right)}^\pi {v_{k}^{Q_2} \left( {{z_u},l,\theta } \right)  {w_{Q_2,k}} \left( {s,{z_u},l,\theta}\right)\!\mathsf{d}\theta \mathsf{d}l}   \Bigg),
\end{split}
\end{equation}
where
\begin{equation}\label{w_Q_2,k}
\begin{split}
    &{w_{Q_2,k}} \left( {s,r} \right) = 1 - {\left( {\frac{m_{Q_2}}{{m_{Q_2}} + s  {\eta _{Q_2}}{\rho_k}{G_{Q_2}}{{\left( {d_{u2U}^2\left( {{z_u},l,\theta } \right) +h_k^2} \right)}^{ \frac{-\alpha _{Q_2}}{2}}}}} \right)^{\frac{m_{Q_2}}{2}}},
\end{split}
\end{equation}

 ${v_k^{Q}\left( {{z_u},l,\theta } \right)}$ and $\varphi_{Q,j,k} \left( {l,r,{z_u}} \right)$ are given in (\ref{vkQ}) and (\ref{phi}), respectively.  lemma that $Q_1$ represents the type of associated UAV, while $Q_2$ represents the type of interfering UAVs. 

\begin{proof}
See Appendix~\ref{app:LT of interference}.
\end{proof}
\end{lemma}

As the distance distributions of tagged UAVs and the association probabilities have been derived, we are ready to calculate the local coverage probability. The definition and derivation of the local coverage probability are given as follows.

\begin{definition}[Local coverage probability]
The local coverage probability ${P^C}\left( {z_u,\gamma} \right)$ is the probability that the SINR of the typical user at distance $z_u$ from the origin is greater than threshold $\gamma$. 
\end{definition}

\begin{theorem}\label{exact theorem}
The exact coverage probability ${P^C}\left( {z_u,\gamma} \right)$ for the typical user is given by,
\begin{sequation}
\begin{split}
\label{exact coverage probability}
    {P^C}  \left( {z_u,\gamma} \right) & = \sum\limits_{k = 1}^K \int_{{h_k}}^{ + \infty } {{f_{{\rm{RLoS}},k}}\left( {r,z_u} \right)} P_{{\rm{LoS}},k}^A\left( {r,z_u} \right)  \sum\limits_{n = 0}^{{m_{{\rm{LoS}}}} - 1} {\left[ {\frac{{{{\left( { - s} \right)}^n}}}{{n!}}\frac{{{\partial ^n}}}{{\partial {s^n}}}{\mathcal{L}_{U_{\rm{LoS},k}}}\left( {s,r,z_u} \right)} \right]}_{s = {\mu _{{\rm{LoS},k}}}\left( {r,\gamma } \right)} \mathsf{d}r \\
    &  + \sum\limits_{k = 1}^K \int_{{h_k}}^{ + \infty } {f_{{\rm{RNLoS}},k}}\left( {r,z_u} \right)P_{{\rm{NLoS}},k}^A\left( {r,z_u} \right) \sum\limits_{n = 0}^{{m_{{\rm{NLoS}}}} - 1} {{{\left[ {\frac{{{{\left( { - s} \right)}^n}}}{{n!}}\frac{{{\partial ^n}}}{{\partial {s^n}}}{\mathcal{L}_{U_{\rm{NLoS},k}}}\left( {s,r,z_u} \right)} \right]}_{s = {\mu _{{\rm{NLoS}}}}\left( {r,\gamma } \right)}}} \mathsf{d}r  ,
\end{split}
\end{sequation}
\begin{equation}
\label{L_U}
    {\mathcal{L}_{U_{Q,k}}}\left( s,r,z_u \right) = \exp \left( { - {\sigma ^2}s} \right){\mathcal{L}_{I_{Q,k}}}\left( s,r,z_u \right),
\end{equation}
\begin{equation}
\label{miu}
    {\mu_{Q,k}}\left( r \right){=}m_Q \gamma \eta _Q^{ - 1}\rho_k^{ - 1}{r^{{\alpha _Q}}},
\end{equation}
$Q = \{ {\rm{LoS}},{\rm{NLoS}}\}$ in (\ref{L_U}) and (\ref{miu}), ${f_{{\rm{R}}Q,k}}\left( {r,z_u} \right)$, ${P_{{\rm{LoS}},k}^A\left( {r,z_u} \right)}$ and ${P_{{\rm{NLoS}},k}^A\left( {r,z_u} \right)}$ are defined in (\ref{sequation: PDF of tagged UAV}), (\ref{PALoSj}) and  (\ref{PANLoSj}).
\begin{proof}
See Appendix~\ref{app:exact theorem}.
\end{proof}
\end{theorem}

Based on the local coverage probability, the definition and derivation of the overall coverage probability are given as follows. 
\begin{definition}[Overall coverage probability]
The overall coverage probability is the average coverage probability of all users. 
\end{definition}
From the definition, the overall coverage probability for the typical user is the normalized expectation of the local coverage probability with regard to $z_u$.
\begin{corollary}\label{overall exact}
The overall exact coverage probability with the \ac{SINR} threshold $\gamma$ is given by,
\begin{equation}
    {P_{\rm{Overall}}^C}\left( \gamma  \right) = \frac{{\int_0^{ + \infty } {\Lambda_u}\left( {{z_u}} \right) {{P^C}\left( {{z_u},\gamma } \right){z_u}\mathsf{d}{z_u}} }}{{\int_0^{ + \infty } {{\Lambda_u}\left( {{z_u}} \right){z_u}\mathsf{d}{z_u}} }}.
\end{equation}

\end{corollary}

As is shown in (\ref{exact coverage probability}), higher-order derivatives of the Laplace transform are needed while deriving the exact coverage probability. Because the computational complexity increases rapidly as the order of the derivative increases, the amount of computation is not acceptable under large shape parameters $m_{\rm{LoS}}$ and $m_{\rm{NLoS}}$. Therefore, we provide an approximate evaluation of the coverage probability using the upper bound of the CDF of the Gamma distribution \cite{approximate_coverage_probability}. 

\begin{theorem}\label{approximate theorem}
The approximate coverage probability ${\widetilde{P}^C}\left( {z_u,\gamma} \right)$ for the typical user is given by,
\begin{sequation}
\begin{split}
\label{approximate coverage probability}
    {\widetilde{P}^C} & \left({z_u}, \gamma \right) = \sum\limits_{k = 1}^K \int_{{h_k}}^{ + \infty } {{f_{{\rm{RLoS}},k}}\left( {r,z_u} \right)} P_{{\rm{LoS}},k}^A\left( {r,z_u} \right) \sum\limits_{n = 1}^{{m_{{\rm{LoS}}}}} {\binom{m_{{\rm{LoS}}}}{n}{{\!\left( { - 1} \right)}^{n + 1}}\!\!{\mathcal{L}_{U_{\rm{LoS},k}}}\!\!\left( {n\,{\omega _{\rm{LoS}}}\,{\mu _{{\rm{LoS},k}}}\left( {r,\gamma } \right)},r,z_u \right)}   \mathsf{d}r \\
    &\!\! + \sum\limits_{k = 1}^K \int_{{h_k}}^{ + \infty } {f_{{\rm{RNLoS}},k}}\left( {r,z_u} \right)P_{{\rm{NLoS}},k}^A\left( {r,z_u} \right) \sum\limits_{n = 1}^{{m_{{\rm{NLoS}}}}}\!\! {\binom{m_{{\rm{NLoS}}}}{n}{{\!\left( \!{ - 1}\! \right)}^{n + 1}}\!\!{\mathcal{L}_{U_{\rm{NLoS},k}}}\!\!\left( {n\,{\omega _{\rm{NLoS}}}\,{\mu _{{\rm{NLoS}}}}\!\left( {r,\gamma } \right)},\!r,\!z_u \right)}  \mathsf{d}r ,
\end{split}
\end{sequation}
where
\begin{equation}
    \omega_Q = \left( {m_Q}! \right)^{-\frac{1}{m_Q}},Q = \{ {\rm{LoS}},{\rm{NLoS}}\},
\end{equation}
${f_{{\rm{R}}Q,k}}\left( {r,z_u} \right)$, ${P_{{\rm{LoS}},k}^A\left( {r,z_u} \right)}$, ${P_{{\rm{NLoS}},k}^A\left( {r,z_u} \right)}$ and ${\mu_{Q,k}}$ are defined in (\ref{sequation: PDF of tagged UAV}), (\ref{PALoSj}), (\ref{PANLoSj}) and (\ref{miu}).
\begin{proof}
See Appendix~\ref{app:approximate theorem}.
\end{proof}
\end{theorem}

The same as the overall exact coverage probability, the overall approximate coverage probability is given in the following corollary. 
\begin{corollary}
The overall approximate coverage probability is given by,
\begin{equation}
\label{overall approximate}
    {\widetilde{P}_{\rm{Overall}}^C}\left( \gamma  \right) = \frac{{\int_0^{ + \infty } {\Lambda_u}\left( {{z_u}} \right) {{\widetilde{P}^C}\left( {{z_u},\gamma } \right){z_u}\mathsf{d}{z_u}} }}{{\int_0^{ + \infty } {{\Lambda_u}\left( {{z_u}} \right){z_u}\mathsf{d}{z_u}} }}.
\end{equation}
\end{corollary}

\section{Numerical Results}
In this section, we compare the coverage performance of different systems and optimize the overall coverage probability by changing the distribution of UAVs in different tiers. Referring to \cite{UAV_SG,al2014optimal,kulkarni2014coverage}, we assume the channel parameters as follows: the LoS and NLoS path-loss exponents are $\alpha_{\rm{LoS}}=2$ and $\alpha_{\rm{NLoS}}=3$, the mean additional gains for LoS and NLoS transmissions are $\eta_{\rm{LoS}}=0\rm{dB}$ and $\eta_{\rm{NLoS}}=-20\rm{dB}$, $m$ parameters of Nakagami-m fading for LoS and NLoS UAVs are $m_{\rm{LoS}}=2$ and $m_{\rm{NLoS}}=1$, the noise power is $\sigma^2 = 10^{-7} \rm{W}$, 
the parameters for the probability of establishing an LoS link in (\ref{PLoSk}) are $a=4.88$ and $b=0.429$. The deterministic parameters of users' distribution in (\ref{Lambda_u}) are $\lambda_u = 10^{-3}\rm{m}^{-2}$ and $\beta_u = 5 \times 10^{-3}$. As a non-homogeneous PPP, the distribution of users can be realized by thinning property  \cite{haenggi2012stochastic} of homogeneous PPP. Finally, we assume three tiers of UAVs are deployed in a small town center square with sides of 5km, at 50,100 and 150 meters height, with the corresponding transmission power 2, 7, and 12\,\rm{dBm}, respectively, and the same value of $\lambda_1 = \lambda_2 = \lambda_3 = 4 \times 10^{-5}\rm{m}^{-2}$.

\begin{table*}[]
\centering
\caption{Table of Parameters}\label{table2}
\begin{tabular}{|c|c|c|c|c|}
\hline
& $h_k$ [m]    & $\beta $   & $P^C_{ \begin{tiny}
 {\rm{Overall}}    \end{tiny} }$  &Density of UAVs \\ \hline
One-tier   & 50           & 3.2 $\times 10^{-3}$               & 0.9026      &$1\,\rm{UAV}/\rm{km}^2$           \\ \hline
One-tier   & 100     & 3.2 $\times 10^{-3}$       & 0.9203       &$1\,\rm{UAV}/\rm{km}^2$          \\ \hline 
One-tier   & 150     & 3.2 $\times 10^{-3}$       & 0.9367       &$1\,\rm{UAV}/\rm{km}^2$          \\ \hline
Three-tier & (50,100,150) & $(4.5,5.8,7.6)$   $\times 10^{-3}$ & 0.9713         &$1\,\rm{UAV}/\rm{km}^2$        \\ \hline
Uniform Distribution   & 100          & 0                                  & 0.3845        &$1\,\rm{UAV}/\rm{km}^2$         \\ \hline
\end{tabular}
\end{table*}

\begin{figure}[h]
 \centering
 {\includegraphics[width=0.65\columnwidth]{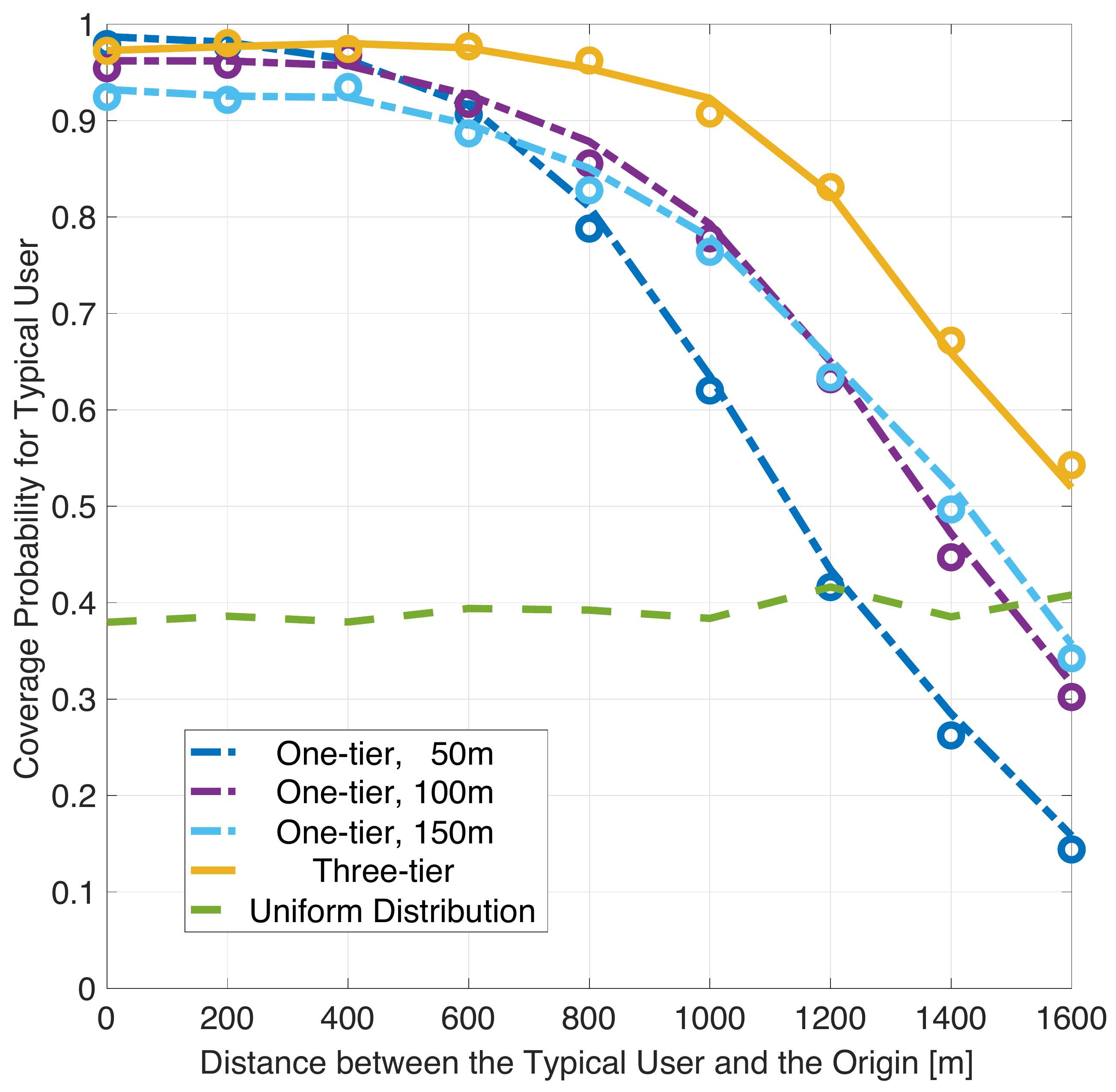}}
 \caption{Local Coverage Probability for the Typical User.}
\label{fig:Local Coverage Probability for Typical User}
\end{figure}

In Fig.~\ref{fig:Local Coverage Probability for Typical User}, a curve of local coverage probability for the typical user as a function of the distance between the origin and the typical user is plotted. We compare the coverage performance of population density-inspired UAV systems (one-tier and three-tier) with that of the uniformly distributed UAV system. All of the above systems have the same UAV deployment density on average as $25 \, \rm{UAVs}/km^2$ (i.e., $\lambda_h = 10^{-6}\rm{m}^{-2}$ and $\beta_h = 0$ for uniform distribution). The distribution parameters $\beta$ of the other three systems are shown in the table, $\lambda =  4 \times 10^{-5}\rm{m}^{-2}$ as mentioned above. 
\par
As shown in Table~\ref{table2}, different $\beta$ are chosen to keep the density of the UAVs as $1\,\rm{UAV}/\rm{km}^2$. Assume the threshold of coverage probability is $\gamma=-15\rm{dB}$. It can be seen that the results of the Monte-Carlo simulation (lines) coincide well with the results of the theoretical analysis (points). With the same number of UAVs, no matter how far away the typical user is from the origin, the performance of the three-tier network is always better than that of the one-tier networks. As shown in Table~\ref{table2}, the three-tier network has significant advantages in terms of overall coverage. At the edge, the advantage of the three-tier network is further expanded, indicating that the lower limit of network coverage or SINR can be guaranteed. In addition, the coverage performance of all three systems is declining from center to edge due to the reduced density of UAV deployment. Compared with the proposed distribution, the uniform distribution system only has a slight advantage in the edge area, but the overall performance is far inferior to the other four systems. The clustering effect brings the non-uniform distribution advantages over the uniform distribution. According to Fig.~\ref{fig:Local Coverage Probability for Typical User}, when the user is close to the center, the coverage probability of the UAV network under resident population density-inspired distribution is significantly greater than that under the uniform distribution. Most users are clustered in the area close to the center, so the proposed distribution has significant advantages in the overall coverage probability.

\begin{figure}[h]
 \centering
 {\includegraphics[width=0.7\columnwidth]{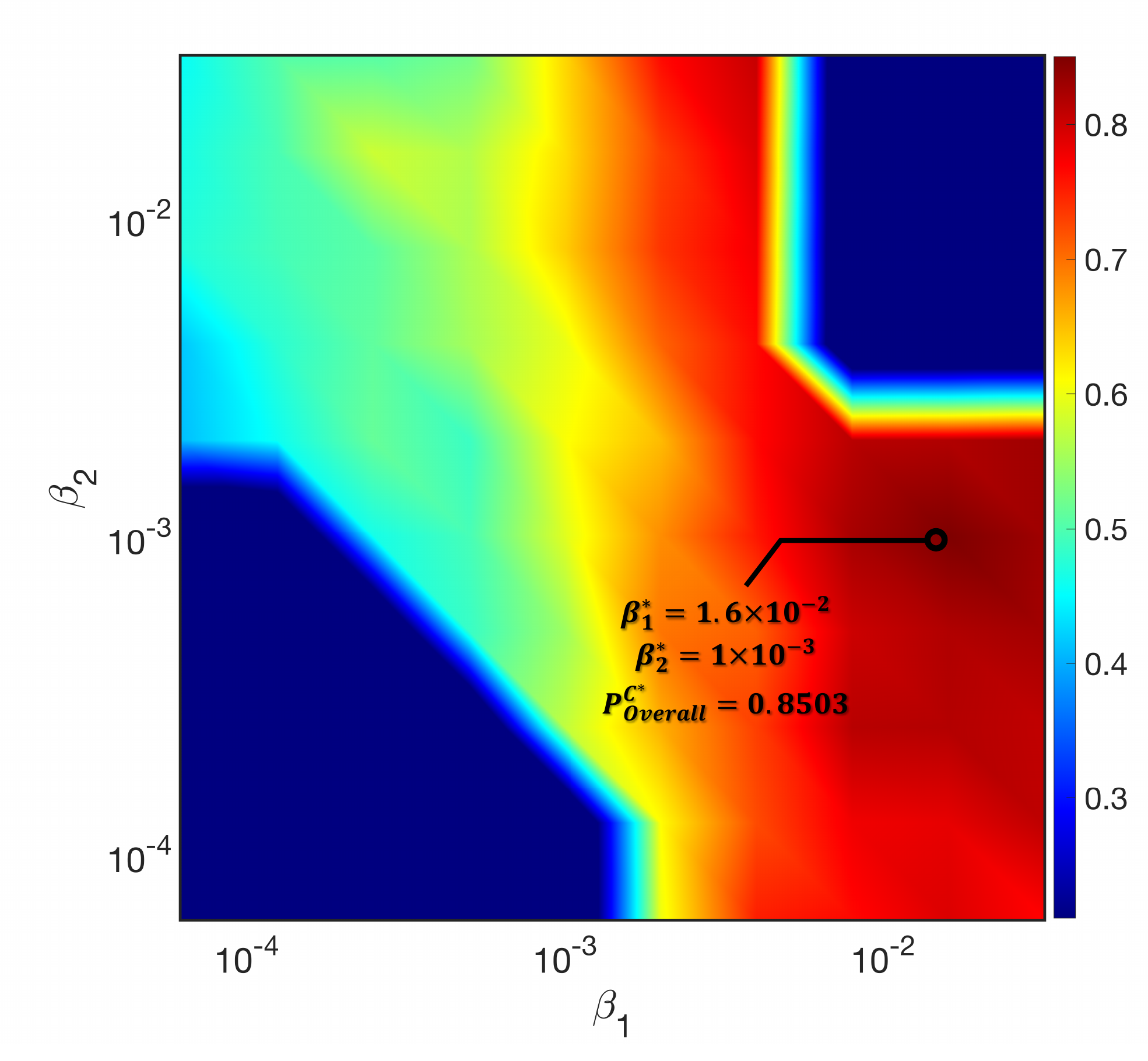}}
 \caption{Overall coverage probability under different UAV distributions.}
\label{fig:Overall coverage probability under different UAV distributions.}
\end{figure}

\begin{figure}[h]
 \centering
 {\includegraphics[width=0.65\columnwidth]{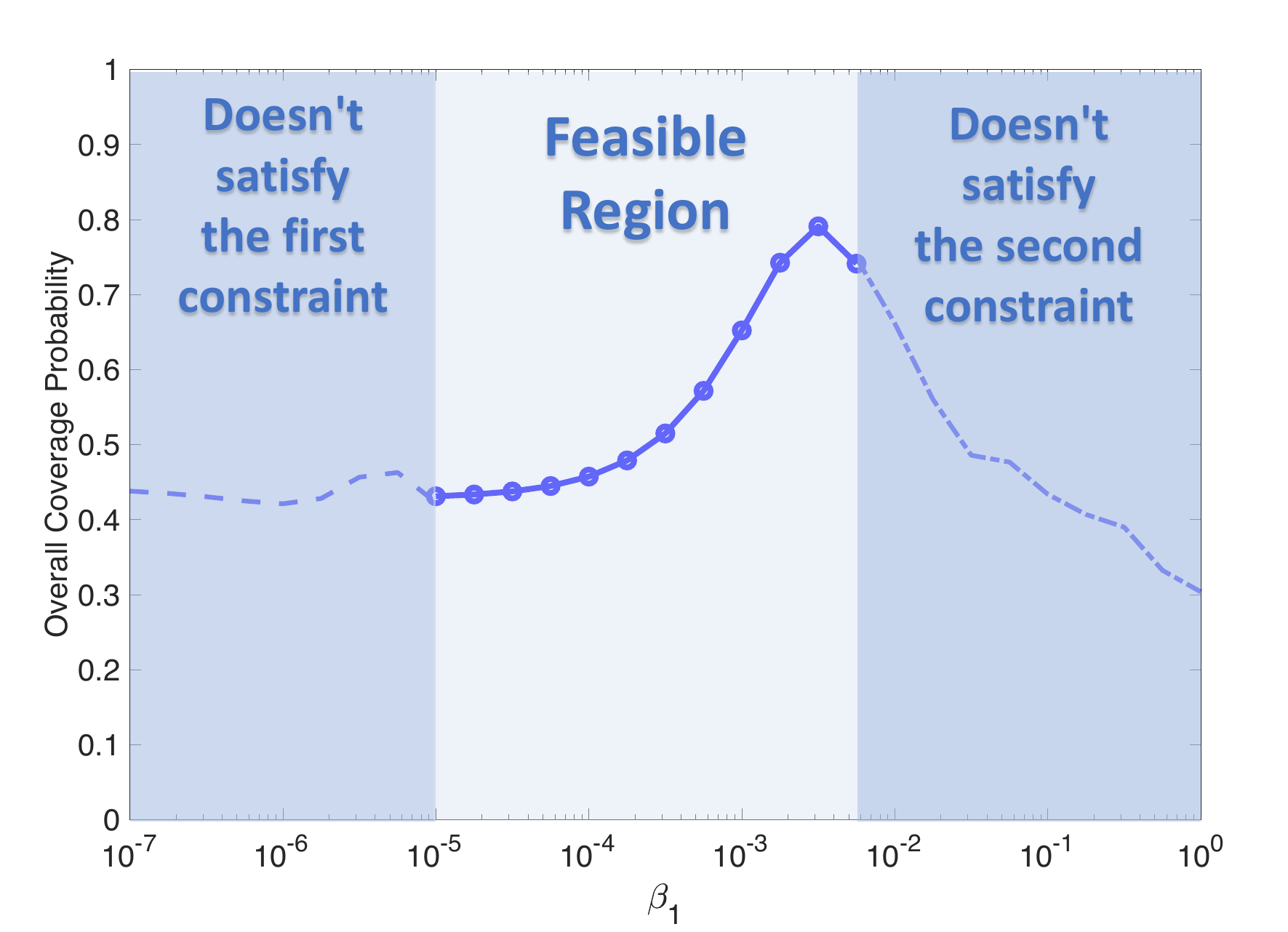}}
 \caption{The influence of the UAV distribution in single tier on the overall coverage probability.}
\label{fig:The influence of the UAV distribution in single tier on the overall coverage probability.}
\end{figure}

We study the following optimization problems and record the results in Fig.~\ref{fig:Overall coverage probability under different UAV distributions.} and Fig.~\ref{fig:The influence of the UAV distribution in single tier on the overall coverage probability.}. We want to maximize the overall coverage probability by changing the distribution of UAVs, under the premise that the total number of UAVs is limited (the first constraint) and the coverage of users in any location is guaranteed (the second constraint). Therefore, the mathematical representation of the optimization problem is as follows,
\begin{equation}
\begin{aligned}
\label{optimization}
    \underset{\beta _1,\beta_2}{\arg \max } \ \ & P^C_{{\rm Overall}} \left (\gamma_1 \right ) \\
    {\rm s.t. \ \ \ \ \ } & \sum_{k=1}^{K}\int_{0}^{+\infty }{\Lambda_{{\rm UAV},k}}\left( {z} \right) \mathrm{d}z\leq N_{\max},\\
    &\mathbb{P}\left [ \rm{SINR}\geq \gamma_2 \right ]\geq 0.95, \ \  \ \ \forall z_u\geq 0,
\end{aligned}
\end{equation}
where the threshold of overall coverage probability is $\gamma_1 = -8\rm{dB}$ and the threshold of local coverage for the typical user is $\gamma_2 = -20\rm{dB}$. The total number of UAVs is limited to $N_{\max}=1000$, that is, the maximum density of UAV is $40\, \rm{UAVs}/\rm{km}^2$. We deploy UAVs at the first the second tiers ($h = 50,100 \, \rm{m}$).
\par
A paradoxical but interesting conclusion in Fig.~\ref{fig:Overall coverage probability under different UAV distributions.} is, the UAV distribution preferred by the system is a non-uniform one influenced by the resident density distribution, but the optimal distribution is not closely related to the residents numerically. The optimal overall coverage probability $P_{\begin{small}\rm{Overall}\end{small}}^{C*}$ and the corresponding $\beta_1^{*}$ and $\beta_2^{*}$ value are marked in the figure. It can be seen that there is a considerable difference among $\beta_1^{*}$, $\beta_2^{*}$ and $\beta_u$. For a large $\beta$, the first constraint cannot be satisfied due to the high density of UAVs. Under this condition, the coverage probability is set to 0, so the dark blue area at the bottom left appears. When $\beta$ increases to $10^{-3}$, there is a large amount of interference near the centre because the system still tends to be uniformly distributed and the density is relatively high. With the increase of $\beta$, the overall coverage probability is improved rapidly. Near the optimal area, the coverage performance of the system is no longer very sensitive to both $\beta_1$ and $\beta_2$. This is interesting because in such a tolerant system, we do not have to select an accurate set of optimal parameters to determine the distribution of UAVs, but only to estimate a range. For a large $\beta$, the small number of drones are almost all concentrated in the central area, making it difficult for users in the edge area to maintain good communication conditions. Therefore, the second constraint cannot be satisfied, and the dark blue area appears at the top right of the image. Finally, with the same number of UAVs as the optimal distribution, the overall coverage probability of the uniform distribution in the same condition ($h=50,100\,\rm{m}$, $\lambda_1 = \lambda_2 =4 \times 10^{-6}\,\rm{m}^{-2}$) is only 0.2883, which is much lower than 0.8503.

Although the optimization problem (\ref{optimization}) has been carefully studied in Fig.~\ref{fig:Overall coverage probability under different UAV distributions.}, it is still necessary to study the behavior of the system hidden in the dark blue area. Fig.~\ref{fig:The influence of the UAV distribution in single tier on the overall coverage probability.} shows the influence of the UAV distribution in a single tier on the overall coverage probability. We observe the special case of $\beta_2=10^{-2}$ in Fig.~\ref{fig:Overall coverage probability under different UAV distributions.} and broaden the range of $\beta_1$. 
\par
First, it is easy to find that the overall coverage probability increases at the beginning and then decreases with the increase of the value of $\beta_1$. This can simply be explained by the fact that too many UAVs will cause too much interference in the central area, while having too few UAVs will make it difficult for the user to find a close UAV to establish an LoS link. When $\beta_1 \leq 10^{-5}$, the number of UAVs in the first tier is much larger than that in the second tier, so the overall coverage probability is stable and tends to be similar to that of uniform distribution in the first tier. When $\beta_1 \geq 5.6 \times 10^{-2}$, the number of UAVs in the first tier is rapidly decreasing, which means there are no available LoS UAVs nearby for some users. It is not hard to predict that the final result will converge to the scenario where only the second tier of UAVs are providing the service.

\section{Further Remarks}
\subsection{Analytic Framework Extension}
This subsection presents how to extend the existing analysis framework to other scenarios and network models. Enhancing the coverage is one of the application scenarios for UAV networks. UAV networks can also relieve the pressure of insufficient channel capacity in town centers. 

\begin{remark}
Based on Shannon's theorem and the definition of coverage probability in (\ref{Definition of coverage probability}), the channel capacity can be expressed as \cite{andrews2011tractable},
\begin{equation}
    \mathbb{P} \left[ B \log_2 \left(1 + \mathrm{SINR} \right) > \mathcal{R} \right] = \mathbb{P} \left[ \mathrm{SINR} > 2^{\frac{\mathcal{R}}{B}} - 1 \right] = \mathbb{P} \left[ \mathrm{SINR} > \widetilde{\gamma} \right],
\end{equation}
where $\widetilde{\gamma}=2^{\frac{\mathcal{R}}{B}} - 1$ is the rate threshold. By replacing $\widetilde{\gamma}$ into SINR threshold $\gamma$, the local probability and overall probability that the channel capacity is greater than the rate threshold can be obtained by ${\widetilde{P}^C} \left({z_u}, \widetilde{\gamma} \right)$ given by (\ref{approximate coverage probability}) and ${\widetilde{P}_{\rm{Overall}}^C}\left( \widetilde{\gamma}  \right)$ given by (\ref{overall approximate}), respectively.
\end{remark}

Next, we study green communications in a small hot spot area centered on a base station. Energy efficiency, the number of bits that can be transmitted per unit of energy consumed, is used as a performance metric for green communication. For convenience, we calculate the energy efficiency as the ratio of the number of bits transmitted per unit time (channel capacity) to the energy consumed per unit time (transmission power). The introduction of the UAV network allows the central base station to reduce its coverage area, thereby reducing transmission power and enhancing energy efficiency. The following remark illustrates how the coverage probability analytic framework can be applied to the above scenario. 

\begin{remark}
From the definition of energy efficiency, it can be calculated by the ratio of channel capacity to transmission power, 
\begin{equation}
    \mathbb{P} \left[ \frac{B}{\rho_k} \log_2 \left(1 + \rm{SINR} \right) > \mathcal{E} \right] = \mathbb{P} \left[ \rm{SINR} > 2^{\frac{\mathcal{E} \rho_k}{B}} - 1 \right] = \mathbb{P} \left[ \rm{SINR} > \widehat{\gamma} \right],
\end{equation}
By replacing $\widehat{\gamma}=2^{\frac{\mathcal{E} \rho_k}{B}} - 1$ into SINR threshold $\gamma$, the local probability and overall probability that the energy efficiency is greater than the rate threshold can be obtained by ${\widetilde{P}^C} \left({z_u}, \widehat{\gamma} \right)$ given by (\ref{approximate coverage probability}) and ${\widetilde{P}_{\rm{Overall}}^C}\left( \widehat{\gamma} \right)$ given by (\ref{overall approximate}), respectively. 
\end{remark}

Furthermore, there is no need for dense deployment of UAVs near the base station in this case. Fortunately, the network model can be easily extended to the above scenario by adjusting the density distribution of UAVs in (\ref{Lambda_U}). Under the premise that the density (whether of the user or the UAV) is only related to the distance to the town center, the analytical framework of this paper is applicable to any distribution model.

\subsection{Distribution Parameter Design}
According to the above theorems, it can be seen that the relationship between the coverage probability and the spatial distribution of UAVs is not straightforward, and obtaining the optimal parameters by optimization tools is challenging. Therefore, the following qualitative criteria for parameters about UAVs' vertical and horizontal distributions are given. Notice that all of the remarks have been verified by simulation.

\begin{remark} 
Remarks on altitudes $h_1, h_2, \dots, h_K$ are given as follow.
\begin{itemize}
    \item In most cases, UAVs have an optimal altitude, and it is better to deploy the UAVs near the optimal altitude. 
    \item While facing a low communication quality, the UAVs are suggested to be distributed at a low altitude so that UAVs are closer to users and users in the LoS region can be covered.
    \item In a good communication environment, by increasing the deployment altitude, more users can establish LoS links with UAVs, therefore, increasing the coverage probability.
\end{itemize}
\end{remark}

\begin{remark}
Remarks on the number of tiers $K$ are given as follows.
\begin{itemize}
    \item With the increase of tiers, more parameters can be optimized so that the coverage performance can be improved to some extent. A network with fewer tiers can be considered a special case of a network with more tiers. However, the improvement in coverage performance is limited when more than three tiers are applied.
    \item Consider a more general case where the receivers (users) can be divided into $M$ classes according to different gain and demodulation capabilities. Multi-tier distribution has significant advantages over single-tier distribution, and the number of deployment tiers $K$ is recommended to be larger than $M$. Assuming that the optimal UAV deployment altitude for the receiver of the $m$-th class is $h_m^*$, the height of UAVs is suggested to satisfy $\min \left\{h_1^*,h_2^*, \dots, h_M^* \right\}<h_k<\max \left\{h_1^*,h_2^*, \dots, h_M^* \right\},\forall k$.
\end{itemize}
\end{remark}

\begin{remark}
Remarks on homogeneity $\beta$ are given as follows.
\begin{itemize}
    \item The value of $\beta$ is related to the strength of interference power relative to noise. When the interference power is significantly stronger than the environmental noise, it is suggested to choose a smaller $\beta$ to make the distribution of UAVs more homogeneous and vice versa. Considering that the value of $\beta_k$ will affect the average number of UAVs in hot areas when $\beta_k$ is changed, $\lambda_k$ is adjusted to keep the average number of UAVs unchanged.
    \item We use the exhaustive search to solve the optimization problem about $\beta$ given in (\ref{optimization}), which results in the calculation complexity increases exponentially with $K$. An improved alternate maximization method can be a substitution for the exhaustive search as described in \cite{mondal2021economic}. The complexity of this method is $\mathcal{O} \left( N K^2\right)$, where $N$ is the preset maximum number of rounds. The set of suboptimal parameters is obtained by optimizing from $\beta_1, \beta_2, \dots$ to $\beta_K$ in order. When optimizing $\beta_k$, the $\beta_k$ is repeatedly reduced by the predefined step size for at most $N$ times, and one of the $\beta$ in the set $\{\beta_1,\beta_2,\dots,\beta_{k-1}\}$ is increased, so that the coverage probability is maximized when the UAV density is unchanged. The optimization of $\beta_k$ ends when the coverage probability no longer increases. 
\end{itemize}
\end{remark}

\begin{table*}[]
\centering
\caption{Optimization of $K$ and $\beta$.}\label{table3}
\begin{tabular}{|c|c|c|c|c|}
\hline
$1\,\rm{UAV}/\rm{km}^2$  & One-tier    & Two-tier   & Three-tier  &  Five-tier   \\ \hline

$h_1=50$m   & N/A       & N/A    & $\beta_1= 4.5 \times 10^{-3}$  &  $\beta_1= 5.4 \times 10^{-3}$  \\   \hline

$h_2=75$m   & N/A     & N/A   & N/A  &  $\beta_2= 6.5 \times 10^{-3}$  \\ \hline 

$h_3=100$m   & N/A     & $\beta_3= 4.2 \times 10^{-3}$       & $\beta_3= 5.8 \times 10^{-3}$   &  $\beta_3= 7.8 \times 10^{-3}$  \\ \hline

$h_4=125$m & N/A &  N/A  & N/A & $\beta_4= 8.2 \times 10^{-3}$  \\ \hline

$h_5=150$m   & $\beta_5= 3.2 \times 10^{-3}$          & $\beta_5= 5.4 \times 10^{-3}$        &  $\beta_5= 7.6 \times 10^{-3}$  &   $\beta_5= 9.8 \times 10^{-3}$   \\ \hline

$P^C_{ \begin{tiny}{\rm{Overall}} \end{tiny} }$   & 0.9367      & 0.9557        & 0.9713   &  0.9786      \\ \hline
\end{tabular}
\end{table*}

The example in Table~\ref{table3} provides further explanation for the above remarks. In Table~\ref{table3}, we compare the coverage performance under different numbers of tiers $K$.  The total density and the density of UAVs in each tier are fixed as $1\,\rm{UAV}/\rm{km}^2$ and $\lambda_k=4 \times 10^{-5}$. The set of homogeneity $\{ \beta_1, \beta_2, \dots, \beta_{K} \}$ is obtained by alternate maximization method. Overall, increasing the number of tiers allows more parameters to be optimized, thus achieving better coverage performance. The UAV deployment in the one-iter network ($\beta_5= 3.2 \times 10^{-3}$) can be regarded as a special case of that of a two-tier network ($\{\beta_3, \beta_5 \}= \{+\infty, 3.2\times 10^{-3} \} $), but it is not optimal. Finally, the gain in coverage probability from deploying more than three tiers of UAV networks is limited.

\section{Conclusion and Future Work}
In this paper, We studied the coverage performance of multi-tier UAV networks in a centralized urban model. We first derived the distance distribution of tagged UAVs and association probability for the selected typical user. Based on this, the analytical expression of downlink coverage probability is given and proved to be consistent with the Monte-Carlo simulation results. As a result, the coverage probability for the typical user and intermediate products are all related to the distance $z_u$. Both the local and total coverage performance are significantly improved by increasing the number of UAV network tiers. The urban population density-inspired model has a huge advantage over the uniform distribution performs. However, too much concentration of UAVs in the central area will bring more noise to the town center and fail to maintain communication for users at the edge. Therefore, how to design the distribution of each tier of UAVs is crucial.
\par
One future research direction is introducing interference and noise mitigation technologies into the framework based on the proposed resident population density-inspired model. In urban areas, the relatively dense deployment of UAVs may cause strong interference. Strong environmental noise in town centers is also one factor limiting the performance of wireless communication. Under the SG framework, orthogonal channel \cite{wang2022ultra} and directional antenna gain \cite{peng2022directional} can be introduced into the system model respectively to reduce interference and noise power. In addition, we model the users as a PPP, which means that the user's movement is undirected and random. Considering that there is a directional flow of people in the town \cite{navigating}, analyzing the coverage probability of the urban system based on SG will be challenging and application-oriented.

\appendices
\section{Proof of Lemma~\ref{CDF of Tagged LoS UAV}}\label{app:CDF of Tagged LoS UAV}
When the distance between the typical user and the origin is fixed, given that the distance between the tagged LoS UAV in tier $k$ and user ${R_{\rm{LoS},k}}$ is a random valuable, the \ac{CDF} of ${R_{\rm{LoS},k}}$ is given by
\begin{equation}
\label{CDF of tagged - 1}
\begin{split}
    {F_{{\rm{RLoS}},k}}\left( {r,z_u} \right) &= {\mathbb{P}}\left[ {{R_{{\rm{LoS}},k}} < r} \right] = 1 - {\mathbb{P}}\left[ {{R_{{\rm{LoS}},k}} > r} \right] \\
    & = 1 - \mathbb{P}\left[ {\mathcal{N}\left( {{\mathcal{A}_k}\left( r \right)} \right) = 0} \right]  \overset{(a)}{=} 1 - \exp \left( { - \int_{{\mathcal{A}_k}\left( r \right)} {{\Lambda _{\rm{UAV},k}}\left( l \right)l\mathsf{d}l\mathsf{d}\theta } } \right),
\end{split}
\end{equation}
where ${\Lambda _{\rm{UAV},k}}\left( l \right)$ is defined in (\ref{Lambda_U}), $\mathcal{N}\left( {{\mathcal{A}_k}\left( r \right)} \right)$ in (\ref{CDF of tagged - 1}) counts the number of the UAVs in region ${{\mathcal{A}_k}\left( r \right)}$, which is a circle  at the height of ${h_k}$ centered directly above the typical user with radius $\sqrt {{r^2} - h_k^2}$, and $(a)$ is given by the property of the general PPP \cite{primer},
\begin{equation}
\begin{split} \label{the property of the general PPP}
    \mathbb{P}\left[ {\mathcal{N}\left( {{{\mathcal{A}_k}\left( r \right)}} \right) = n} \right] = & \exp \left( { - \int_{{{\mathcal{A}_k}\left( r \right)}} {{\Lambda_{\rm{UAV},k}}\left( l \right)l\,\mathsf{d}l\mathsf{d}\theta} } \right)  \frac{{\exp {{\left( { - \int_{{{\mathcal{A}_k}\left( r \right)}} {{\Lambda_{\rm{UAV},k}}\left( l \right)l\,\mathsf{d}l\mathsf{d}\theta} } \right)}^n}}}{{n!}}.
\end{split}
\end{equation}

\begin{figure}[h]
 \centering
 {\includegraphics[width=0.6\columnwidth]{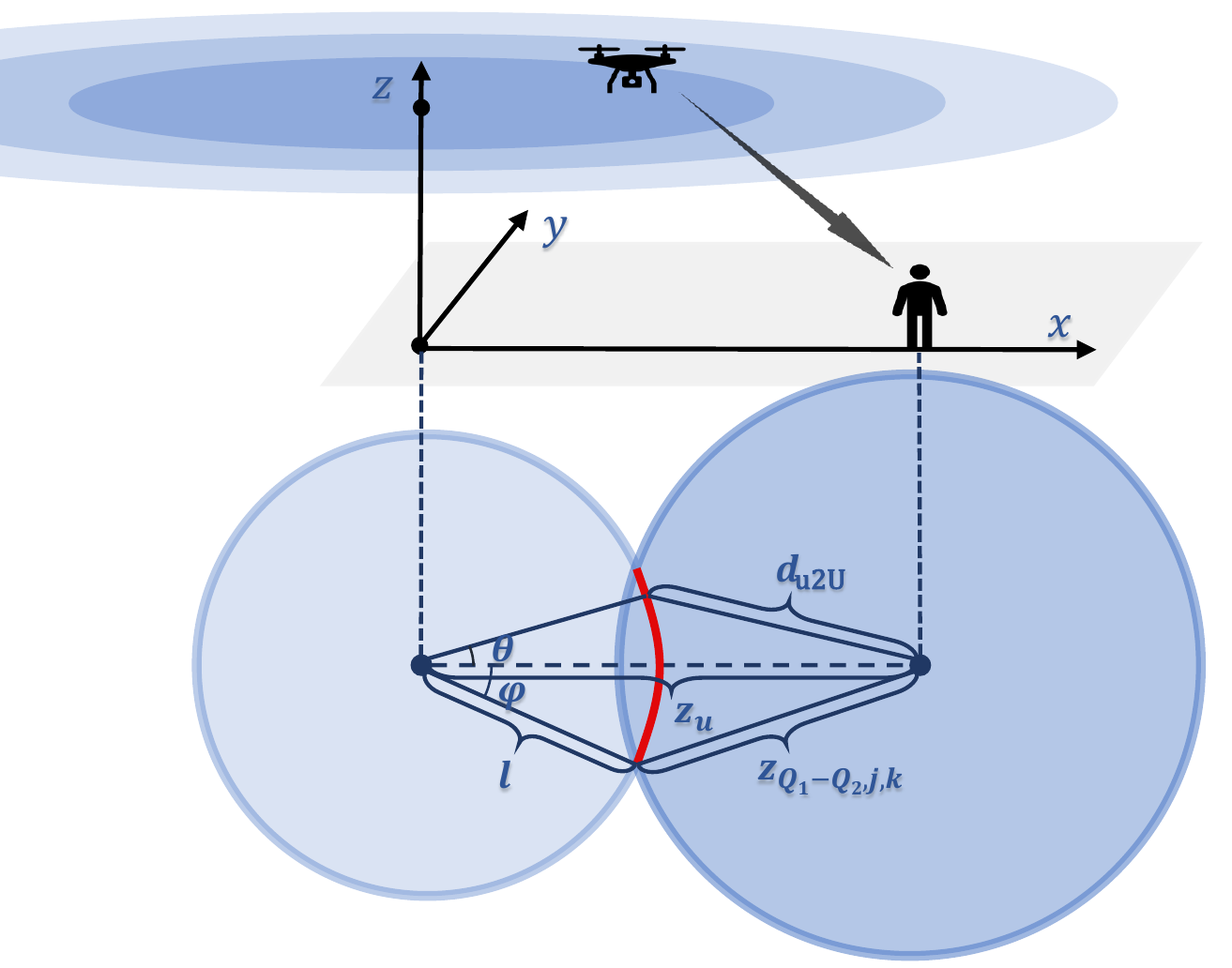}}
 \caption{Vertical Viewed System Schematic Figure.}
\label{fig:Vertical Viewed System Schematic Figure}
\end{figure}

where $z_u$ is the horizontal distance from the typical user to the origin, $\lambda_u$ determines the total density of the plane, $\beta_u$ is a measure of homogeneity.

To integrate formulation in (\ref{CDF of tagged - 1}) over the region of ${{\mathcal{A}_k}}$, the area is divided into infinite concentric circular arcs centered at the point which is directly above the origin at the height of $h_k$. When the radius $l$ of the concentric circular arc is fixed, the density function ${\Lambda _{\rm{UAV},k}}$ is a constant, the coordinates of the points on the arc can be uniquely represented by $\theta$. The bold part of the bottom half of the Fig.~\ref{fig:Vertical Viewed System Schematic Figure} is one of the concentric arcs. The upper bound of $\theta$ can be obtained from the geometric relations in the Fig.~\ref{fig:Vertical Viewed System Schematic Figure}, denoted as $\varphi_{\rm{LoS-LoS}}$, defined in (\ref{phi}), and the lower bound of $\theta$ is $-\varphi_{\rm{LoS-LoS}}$ because of the symmetry of the circle. Furthermore, the horizontal distance ${d_{\rm{u2U}}}\left( {{z_u},l,\theta}\right)$ between the typical user and the point on the arc is defined in (\ref{du2U}), which can also be obtained from simple  geometrical relationships. Hence, we have the following equation
\begin{equation}
\label{CDF of tagged - 2}
\begin{split}
    &\int_{{\mathcal{A}_k}\left( r \right)} {{\Lambda _{\rm{UAV},k}}\left( l \right)l \mathsf{d}l \mathsf{d}\theta }  = \int_{{z_u} - \sqrt {{r^2} - h_k^2} }^{{z_u} + \sqrt {{r^2} - h_k^2} } \int_{ - \varphi_{\rm{LoS-LoS}} \left( {l,r,{z_u}} \right)}^{\varphi_{\rm{LoS-LoS}} \left( {l,r,{z_u}} \right)} {v_k^{{\rm{LoS}}}\left( {{z_u},l,\theta } \right) \mathsf{d}\theta \mathsf{d}l} ,
\end{split}
\end{equation}
where ${{\Lambda _{\rm{UAV},k}}\left( l \right)}$ and ${v_k^{Q}\left( {{z_u},l,\theta } \right)}$ are defined in \ref{Lambda_U}) and (\ref{vkQ}), respectively. It is important to note that ${l}$ may be negative when the value of ${z_{{\rm{LoS - LoS}},j,k}}$ is greater than the horizontal distance ${z_u}$. Therefore, ${|l|}$ is used in the outer integral.


\section{Proof of Lemma~\ref{PDF of Tagged UAV}}\label{app:PDF of Tagged UAV}
As in Lemma~\ref{PDF of Tagged UAV}, $Q$ is used to represent the type of tagged UAVs, i.e., $Q$ is replaced with LoS when an LoS UAV is tagged or NLoS otherwise. By taking the derivative of ${F_{{\rm{R}Q},k}}\left( {r,z_u} \right)$, the distribution of the nearest \acp{UAV} in tier $k$ with a distance $r$ from the user is obtained,  which is denoted as ${f_{{\rm{R}Q},k}}\left( {r,z_u} \right)$,
\begin{equation}
\begin{split}
\label{fRLoS1}
    &{f_{{\rm{R}Q},k}}\left( {r,z_u} \right) = \frac{\partial }{{\partial r}}{F_{{\rm{R}Q},k}}\left( {r,z_u} \right) \\
    & = \frac{\partial }{{\partial r}}\Bigg( 1 -  \exp  \Bigg( -  \int_{{z_u} - \sqrt {{r^2} - h_k^2} }^{{z_u} + \sqrt {{r^2} - h_k^2} }  \int_{ - \varphi_{Q-Q} \left( {l,r,{z_u}} \right)}^{\varphi_{Q-Q} \left( {l,r,{z_u}} \right)}  {v_k^{Q} \left( {{z_u},l,\theta } \right)\mathsf{d}\theta \mathsf{d}l}  \Bigg)  \Bigg)\\
    &\overset{(a)}{=} \exp \Bigg(  - \int_{{z_u} - \sqrt {{r^2} - h_k^2} }^{{z_u} + \sqrt {{r^2} - h_k^2} } \int_{ - \varphi_{Q-Q} \left( {l,r,{z_u}} \right)}^{\varphi_{Q-Q} \left( {l,r,{z_u}} \right)} {v_k^{Q}\left( {{z_u},l,\theta }  \right)\mathsf{d}\theta \mathsf{d}l} \Bigg) \\
    & \times \Bigg(\int_{{z_u} - \sqrt {{r^2} - h_k^2} }^{{z_u} + \sqrt {{r^2} - h_k^2} } \underbrace{\frac{\partial }{{\partial r}}f_{\rm{in},k}\left( {l,{z_u},r,\theta } \right)}_{\begin{small}\rm{The\ derivative\ of\ the\ integrand}\end{small}} \mathsf{d}l \\
    & + \underbrace{\frac{{\partial \left( {{z_u} + \sqrt {{r^2} - h_k^2} } \right)}}{{\partial r}}f_{\rm{in},k}\left( {{z_u} + \sqrt {{r^2} - h_k^2} ,{z_u},r,\theta } \right)}_{\rm{The\ derivative\ of\ the\ integral\ upper\ bound}} \\
    & - \underbrace{\frac{{\partial \left( {{z_u} - \sqrt {{r^2} - h_k^2} } \right)}}{{\partial r}}f_{\rm{in},k}\left( {{z_u} - \sqrt {{r^2} - h_k^2} ,{z_u},r,\theta } \right)}_{\rm{The\ derivative\ of\ the\ integral\ upper\ bound}} \Bigg),
\end{split}
\end{equation}

where ${v_k^{Q}\left( {{z_u},l,\theta } \right)}$ and $\varphi_{Q-Q} \left( {l,r,{z_u}} \right)$ are defined in (\ref{vkQ})  and (\ref{phi}), respectively, and $(a)$ follows Leibnitz's rule, \\ $f_{\rm{in},k}\left( {l,{z_u},r,\theta } \right)$ is the integrand of the outer integral, given by
\begin{equation}
\begin{split}
    &f_{\rm{in},k}\left( {l,{z_u},r,\theta } \right) =  \int_{ - \varphi_{Q-Q} \left( {l,r,{z_u}} \right)}^{\varphi_{Q-Q} \left( {l,r,{z_u}} \right)} {v_k^{Q}\left( {{z_u},l,\theta } \right) \mathsf{d}\theta }.
\end{split}
\end{equation}
\par

For the derivative of the integral upper bound in (\ref{fRLoS1}),
\begin{equation}
\begin{split}
\label{fRLoS2}
    &\frac{{\partial \left( {{z_u} + \sqrt {{r^2} - h_k^2} } \right)}}{{\partial r}}f_{\rm{in},k}\left( {{z_u} + \sqrt {{r^2} - h_k^2} ,{z_u},r,\theta } \right)\\
    & = \frac{{r}}{{\sqrt {{r^2} - h_k^2} }} \int_{ - \varphi_{Q-Q} \left( {{z_u} + \sqrt {{r^2} - h_k^2} ,r,{z_u}} \right)}^{\varphi_{Q-Q} \left( {{z_u} + \sqrt {{r^2} - h_k^2} ,r,{z_u}} \right)} {v_k^{Q} \left( {{z_u},{z_u} + \sqrt {{r^2} - h_k^2} ,\theta } \right) \mathsf{d}\theta }. 
\end{split}
\end{equation}

The derivative of the integral lower bound is similar to that of (\ref{fRLoS2}), therefore omitted here. 


For the derivative of the intergrad, 
\begin{equation}
\label{fRLoS4}
\begin{split}
    &\frac{\partial }{{\partial r}}f_{\rm{in},k}\left( {l,{z_u},r,\theta } \right) = \frac{\partial }{{\partial r}}\int_{ - \varphi_{Q-Q} \left( {l,r,{z_u}} \right)}^{\varphi_{Q-Q} \left( {l,r,{z_u}} \right)} {v_k^{Q} \left( {{z_u},l,\theta } \right) \mathsf{d}\theta } \\
    &\overset{(a)}{=} v_k^{Q} \left( {{z_u},l,\varphi_{Q-Q} } \right) \frac{{\partial \varphi_{Q-Q} \left( {l,r,{z_u}} \right)}}{{\partial r}} - v_k^{Q} \left( {{z_u},l, - \varphi_{Q-Q} } \right) \frac{{\partial \left( { - \varphi_{Q-Q} \left( {l,r,{z_u}} \right)} \right)}}{{\partial r}}\Big)\\
    &\overset{(b)}{=} 2 v_k^{Q} \left( {{z_u},l,\varphi_{Q-Q} } \right)\frac{{\partial \varphi_{Q-Q} \left( {l,r,{z_u}} \right)}}{{\partial r}}\\
    &\overset{(c)}{=} \mathbbm{1}\left( {r > {h_k}} \right) \ \frac{{4 \, r \,v_k^{Q} \left( {{z_u},l,\varphi_{Q-Q} } \right)}}{{\sqrt {4 \, {l^2} \, z_u^2 - {{\left( {{l^2} + z_u^2 - {r^2} + h_k^2} \right)}^2}} }},
\end{split}
\end{equation}
where $(a)$ follows Leibniz's rule for internal integral, and the expression in $(b)$ is simplified by the fact ${d_{{\rm{u2U}}}}\left( {{z_u},l, - \varphi_{Q-Q} } \right) = {d_{{\rm{u2U}}}}\left( {{z_u},l,\varphi_{Q-Q} } \right)$, which can be easily obtained by (\ref{phi}), \\ $\mathbbm{1}\left( {r > {h_k}} \right)$ is the indicator function defined in (\ref{PDF of Tagged UAV}), and $(c)$ is obtained by substitute $\frac{{\partial \varphi_{Q-Q} \left( {l,r,{z_u}} \right)}}{{\partial r}}$, which is given by, 
\begin{equation}
\begin{split}
    &\frac{{\partial \varphi_{Q-Q} \left( {l,r,{z_u}} \right)}}{{\partial r}}  = \frac{\partial }{{\partial u}}\arccos \left( {\frac{{{l^2} + z_u^2 - {u^2}}}{{2 \, l \, {z_u}}}} \right) \ \frac{\partial }{{\partial v}}\sqrt {{v^2} - h_k^2}  \ \frac{{\partial v}}{{\partial r}}\\
    & = \frac{{2u}}{{\sqrt {4 \, {l^2} \, z_u^2 - {{\left( {{l^2} + z_u^2 - {u^2}} \right)}^2}} }}  \frac{{2v}}{{2\sqrt {{v^2} - h_k^2} }} \cdot \mathbbm{1}\left( {{{\left( {\frac{{{\rho_k}}}{{{\rho_k}}}} \right)}^{\frac{1}{{{\alpha _{{\rm{LoS}}}}}}}}r > {h_k}} \right)\\
    & = \mathbbm{1}\left( {r > {h_k}} \right) \ \frac{{2r}}{{\sqrt {4 \, {l^2} \, z_u^2 - {{\left( {{l^2} + z_u^2 - {r^2} + h_k^2} \right)}^2}} }},
\end{split}
\end{equation}
where $u = {z_{{\rm{LoS - LoS}},j,k}}\left( r \right)\left| {_{j = k}} \right. = \mathbbm{1}\left( {r > {h_k}} \right) \, \sqrt {{r^2} - h_k^2} $, and $v = {d_{{\rm{LoS - LoS}},j,k}}\left( r \right)\left| {_{j = k}} \right. = \max \left\{ {{h_k},r} \right\}$. Substitute (\ref{fRLoS2}) and (\ref{fRLoS4}) into (\ref{fRLoS1}), the final result is derived.

\begin{figure*}[h]
	\centering
	\includegraphics[width=0.8 \linewidth]{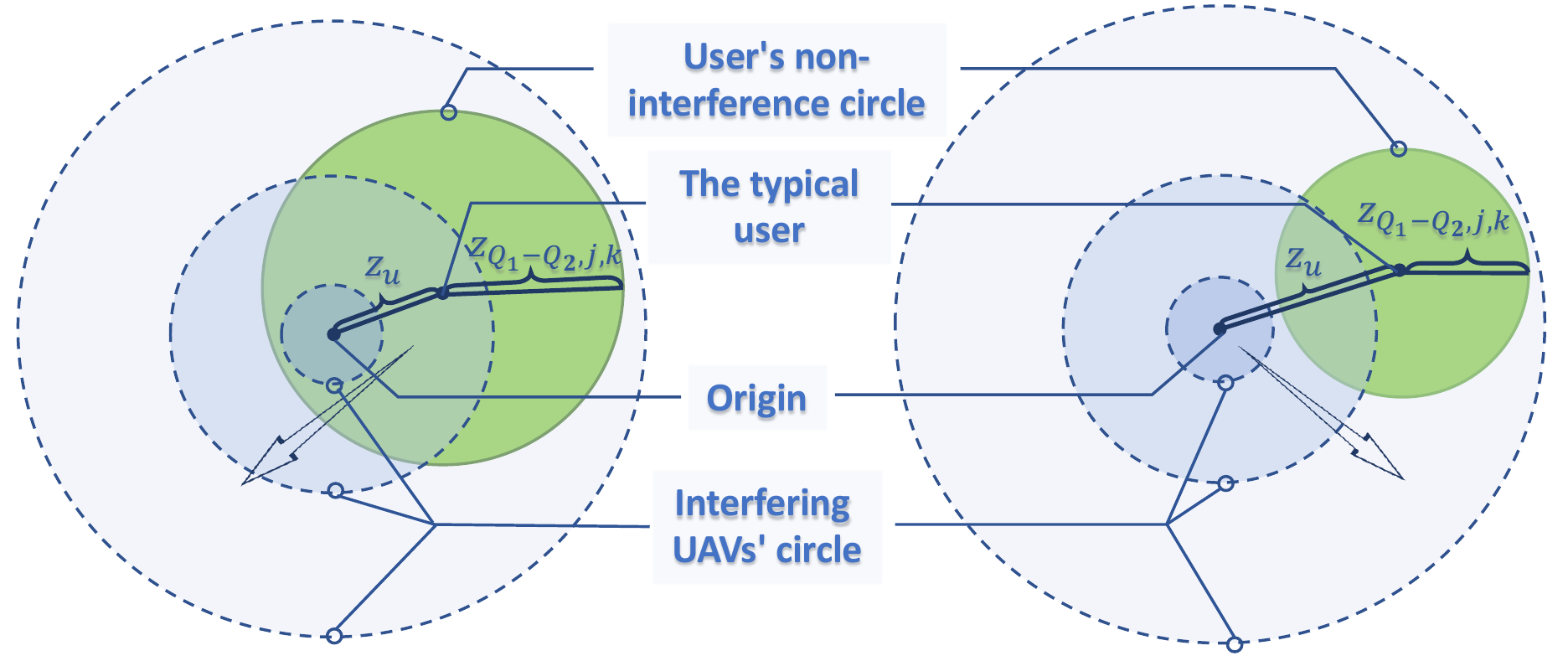}
	\caption{Two relationships between user's non-interference circle and interfering UAVs' circle.}
	\label{fig:Two relationships between user's non-interference circle and interfering UAVs' circle}
\end{figure*}

\section{Proof of Lemma~\ref{Association of LoS UAV}}\label{app:Association of LoS UAV}
When the distance between the typical user and the origin is fixed, the probability that the typical user is associated with the tagged LoS UAV in tier $j$ is equal to the probability that the average received power of other $2K-1$ tagged UAVs is lower than it, where $K$ is the number of tiers. Using the solution of lemma~\ref{nearest interfering LoS}, we have
\begin{equation}
\label{Association of LoS UAV - 1}
\begin{split}
    &P_{{\rm{LoS}},j}^A({r,z_u}) = {\prod\limits_{k = 1,j \ne k}^K{\mathbb{P}}\left[ {{R_{{\rm{LoS}},k}} > {d_{{\rm{LoS - LoS}},j,k}}\left( r \right)} \right]} \times  {\prod\limits_{k = 1}^K  {\mathbb{P}}\left[ {{R_{{\rm{NLoS},k}}} > {d_{{\rm{LoS - NLoS}},j,k}}\left( r \right)} \right] }  \\
    & = {\prod\limits_{k = 1,j \ne k}^K{\mathbb{P}}\left[ {\mathcal{N}\left( {{\mathcal{A}_{\rm{LoS},k}}\left( r \right)} \right) = 0} \right]} \times {\prod\limits_{k = 1}^K  {\mathbb{P}}\left[ {\mathcal{N}\left( {{\mathcal{A}_{\rm{NLoS},k}}\left( r \right)} \right) = 0} \right] }\\
    &\overset{(a)}{=} {\prod\limits_{k = 1,j \ne k}^K \exp \left( { - \int_{{\mathcal{A}_{\rm{LoS},k}}\left( r \right)} {  {v_k^{\rm{LoS}} \left( {{z_u},l,\theta } \right)}\mathsf{d}l\,\mathsf{d}\theta } } \right)} \times  {\prod\limits_{k = 1}^K \exp \left( { - \int_{{\mathcal{A}_{\rm{NLoS},k}}\left( r \right)} {    {v_k^{\rm{NLoS}}\left( {{z_u},l,\theta } \right)} \mathsf{d}l\,\mathsf{d}\theta } } \right)},
\end{split}
\end{equation}
where ${R_{Q,k}}$ is the distance between the tagged UAV in tier $k$ and typical user, ${v_k^{Q}\left( {{z_u},l,\theta } \right)}$ is defined in  (\ref{vkQ}), $\mathcal{N}\left( {{\mathcal{A}_{Q,k}}\left( r \right)} \right)$ counts the number of the UAVs in region ${{\mathcal{A}_{Q,k}}\left( r \right)}$, which is a circle at the height of ${h_k}$ centered directly above the typical user with radius ${z_{\rm{LoS}-Q,j,k}}\left( r \right)$, $Q = \left\{ \rm{LoS},\rm{NLoS} \right\}$, and $(b)$ is given by the property of the general PPP in (\ref{the property of the general PPP}). The following two equations can be obtained in a similar way to (\ref{CDF of tagged - 2}), 
\begin{equation}
\label{Association of LoS UAV - 2}
\begin{split}
    &\int_{{\mathcal{A}_{Q,k}}\left( r \right)} {     {v_k^{Q}\left( {{z_u},l,\theta } \right)} \mathsf{d}l\,\mathsf{d}\theta }  = \int_{{z_u} - {z_{{Q - Q},j,k}}}^{{z_u} + {z_{Q - Q,j,k}}} \int_{ - \varphi{_{Q - Q}}\left(l,r,z_u\right)}^{\varphi{_{Q - Q}}\left(l,r,z_u\right)} {v_k^{Q}({z_u,l,\theta }})\mathsf{d}\theta \mathsf{d}l ,
\end{split}
\end{equation}
where $Q = \left\{ \rm{LoS},\rm{NLoS} \right\}$. The final result is derived by substituting (\ref{Association of LoS UAV - 2}) into (\ref{Association of LoS UAV - 1}).

\section{Proof of Lemma~\ref{LT of interference}}\label{app:LT of interference}
For LoS associated UAV in tier $j$, the Laplace transform of the interference power can be expressed as, 
\begin{equation}
\begin{split}
    &{\mathcal{L}_{I_{\rm{LoS},j}}}\left( {s,r,z_u} \right) \overset{(a)}{=} \mathbb{E}_{I_{\rm{LoS},j}} \left[ {{e^{ - s{I_{\rm{LoS},j}}}}} \right]\\
    & = {\mathbb{E}_{{\Phi},G}}\Bigg[ \exp \Bigg(  - s\sum\limits_{k=1}^K \Bigg( \sum\limits_{x \in {\Phi _{{\rm{LoS}},k}} \backslash {x_o}} {{\eta _{{\rm{LoS}}}}{\rho_k}{G_{{\rm{LoS}}}}{r^{ - {\alpha _{{\rm{LoS}}}}}}}   + \sum\limits_{x \in {\Phi _{{\rm{NLoS}},k}}} {{\eta _{{\rm{NLoS}}}}{\rho_k}{G_{{\rm{NLoS}}}}{r^{ - {\alpha _{{\rm{NLoS}}}}}}}  \Bigg)  \Bigg) \Bigg]\\
    & \overset{(b)}{=} \prod\limits_{k=1}^K {\mathbb{E}_{{\Phi _{{\rm{LoS}},k}}}}\Bigg[ \prod\limits_{x \in {\Phi _{{\rm{LoS}},k}} \backslash {x_o}} {{\mathbb{E}_{{G_{{\rm{LoS}}}}}}\left[ {\exp \left( { - s \, {\eta _{{\rm{LoS}}}}{\rho_k}{G_{{\rm{LoS}}}}{r^{ - {\alpha _{{\rm{LoS}}}}}}} \right)} \right]}  \Bigg] \\
    & \times \prod\limits_{k=1}^K {\mathbb{E}_{{\Phi _{{\rm{NLoS}},k}}}}\Bigg[ \prod\limits_{x \in {\Phi _{{\rm{NLoS}},k}}} {{\mathbb{E}_{{G_{{\rm{NLoS}}}}}}\left[ {\exp \left( { - s \, {\eta _{{\rm{NLoS}}}}{\rho_k}{G_{{\rm{NLoS}}}}{r^{ - {\alpha _{{\rm{NLoS}}}}}}} \right)} \right]}  \Bigg],
\end{split}
\end{equation}
where $(a)$ follows the definition of Laplace transform, $\\{{\Phi _{{\rm{LoS}},k}}\backslash {\Phi _o}}$ are all of the LoS UAVs in tier $k$ except for the associated one, and $(b)$ follows the independence of the point process and the small scale fading, $(c)$ is obtained. For Laplace transform of the interference caused by LoS UAVs in tier $k$, 

\begin{table*}
\begin{sequation}
\label{LT of LoS associated and LoS interfering}
\begin{split}
     \qquad &{{\mathbb{E}_{{\Phi _{{\rm{LoS}},k}}}}\left[ {\prod\limits_{x \in {\Phi _{{\rm{LoS}},k}} \backslash {x_o}} {{\mathbb{E}_{{G_{{\rm{LoS}}}}}}\left[ {\exp \left( { - s \, {\eta _{{\rm{LoS}}}}{\rho_k}{G_{{\rm{LoS}}}}{r^{ - {\alpha _{{\rm{LoS}}}}}}} \right)} \right]} } \right]}\\
     &\overset{(a)}{=} \exp ( - \int_{\mathcal{A}_{\rm{LoS},k}^C(r)}  v_k^{\rm{LoS}} \left(  {z_u},l,\theta  \right) {\left( {1 - {\mathbb{E}_{{G_{{\rm{LoS}}}}}}\left[ {\exp \left( { - s \, {\eta _{{\rm{LoS}}}}{\rho_k}{G_{{\rm{LoS}}}}{\left({d_{u2U}^2\left( {{z_u},l,\theta } \right) + h_k^2}\right)^{ - {\alpha _{{\rm{LoS}}}}/2}}} \right)} \right]} \right)} \,\mathsf{d}\theta \mathsf{d}l) \\
    &= \exp \Bigg( - \int_{\mathcal{A}_{\rm{LoS},k}^C(r)} v_k^{\rm{LoS}} \left( {{z_u},l,\theta }  \right)   {\left( {1 - {{\left( {\frac{{{m_{{\rm{LoS}}}}}}{{{m_{{\rm{LoS}}}} + s \, {\eta _{{\rm{LoS}}}}{\rho_k}{G_{{\rm{LoS}}}}{\left({d_{u2U}^2\left( {{z_u},l,\theta } \right) + h_k^2}\right)^{ - {\alpha _{\rm{LoS}}}/2}}}}} \right)}^{m_{{\rm{LoS}}}}}} \right)} l\,\mathsf{d}\theta \mathsf{d}l\Bigg)\\
    &\overset{(b)}{=} \exp \Bigg(  - \underbrace{  \int_0^{\max \left\{ {0,{z_u} - {z_{\rm{LoS-LoS},j,k}}\left( r \right)} \right\}} \int_{ - \pi }^\pi  {v_k^{{\rm{LoS}}}\left( {{z_u},l,\theta }\right){w_{{{\rm{LoS}}},k}}\left( s,{z_u},l,\theta \right)\mathsf{d}\theta \mathsf{d}l} }_{\rm{User's\ non-interference \ circle\ separates\ from\ interfering \ UAVs'\ circle}} \Bigg) \\
    & \times \exp \Bigg( - \underbrace{  \int_{{z_u} + {z_{\rm{LoS-LoS},j,k}}\left( r \right)}^{ + \infty }  \int_{ - \pi }^\pi  {v_k^{{\rm{LoS}}}\left( {{z_u},l,\theta }  \right){w_{{\rm{LoS}},k}}\left( s,{z_u},l,\theta \right)\mathsf{d}\theta \mathsf{d}l}}_{\rm{User's\ non-interference\ circle\ contained\ by\ interfering \ UAVs'\ circle}} \Bigg)\\
    & \times \exp \Bigg(- \underbrace{ 2\int_{{{z_u} - {z_{\rm{LoS-LoS},j,k}}\left( r \right)}}^{{z_u} + {z_{\rm{LoS-LoS},j,k}}\left( r \right)} \int_{\varphi_{\rm{LoS-LoS},j,k} \left( {l,r,{z_u}} \right)}^\pi  {v_k^{{\rm{LoS}}}\left( {{z_u},l,\theta }\right){w_{{\rm{LoS}},k}}\left( s,{z_u},l,\theta  \right)\mathsf{d}\theta \mathsf{d}l} }_{\rm{User's\ non-interference\ circle\ and\ interfering \ UAVs'\ circle\ intersect }} \Bigg) \\
\end{split}
\end{sequation}
\end{table*}

In equation (\ref{LT of LoS associated and LoS interfering}) shown at the top of the next page, $(a)$ follows the PGFL of inhomogeneous PPP  \cite{primer},  $\mathcal{A}_{\rm{LoS},k}^C(r)$ is the complement of $\mathcal{A}_{\rm{LoS},k}(r)$ in the two dimensional plane at the height of ${h_k}$, the definition of   $\mathcal{A}_{\rm{LoS},k}(r)$ is described in Appendix~\ref{app:CDF of Tagged LoS UAV},  ${v_k^{Q}\left( {z_u},l,\theta \right)}$ and $\varphi_{Q-Q} \left( {l,r,{z_u}} \right)$ are defined in (\ref{vkQ}) and (\ref{phi}) respectively, ${w_{{\rm{LoS,}}k}}\left( {s,r} \right)$ defined in (\ref{w_Q_2,k}) is used to simplify the expression in $(b)$.

As is shown in Fig.~\ref{fig:Two relationships between user's non-interference circle and interfering UAVs' circle}, there should be non-interfering UAVs inside the green circle centered at the typical user, the green circle is called the user's non-interference circle. The difference between the left and right images is whether the origin is included by user's non-interference circle. For a fixed radius $l$, the circles centered at the origin is used to cover the possible locations of interfering UAVs with horizontal distance $l$ to the origin, called the interfering UAV's circle. These two circles may be separated or intersected, and sometimes one circle may contain another, shown in step $(b)$ of (\ref{LT of LoS associated and LoS interfering}). The Laplace transform of the interference in other conditions is similar to the process in (\ref{LT of LoS associated and LoS interfering}), therefore omitted here.

\section{Proof of Theorem~\ref{exact theorem}}\label{app:exact theorem}
By the definition of coverage probability in (\ref{Definition of coverage probability}), SINR becomes a deterministic expression only when: (\romannumeral1) the tier where the associated UAV is located; (\romannumeral2) LoS or NLoS link constructed by the typical user and associated UAV; (\romannumeral3) the distance between the typical user and the origin; (\romannumeral4) the Euclidean distance between the typical user and the associated UAV. Therefore, the coverage probability of the typical user is given by (\ref{exact coverage probability_1}) at the top of next page.
\begin{table*}
\begin{equation}
\begin{split}
\label{exact coverage probability_1}
    &{P^C}\left( {{z_u},\gamma } \right) =\sum\limits_{k = 1}^K {{\mathbb{E}_{r,I}}\left[ {P_{{\rm{LoS}},k}^A\left( {r,z_u} \right)\mathbb{P}\left[ {\frac{{{\eta _{{\rm{LoS}}}}{\rho_k}{G_{{\rm{LoS}}}}{r^{ - {\alpha _{{\rm{LoS}}}}}}}}{{I_{\rm{LoS},k}\left( {r,z_u} \right) + {\sigma ^2}}} > \gamma } \right]} \right]} \\
    & + \sum\limits_{k = 1}^K {{\mathbb{E}_{r,I}}\left[ {P_{{\rm{NLoS}},k}^A\left( {r,z_u} \right)\mathbb{P}\left[ {\frac{{{\eta _{{\rm{NLoS}}}}{\rho_k}{G_{{\rm{NLoS}}}}{r^{ - {\alpha _{{\rm{NLoS}}}}}}}}{{I_{\rm{NLoS},k}\left( {r,z_u} \right) + {\sigma ^2}}} > \gamma } \right]} \right]}\\
    &\overset{(a)}{=} \sum\limits_{k = 1}^K {{\mathbb{E}_{r,U}}\left[ {P_{{\rm{LoS}},k}^A\left( {r,z_u} \right)\mathbb{P}\left[ {{G_{{\rm{LoS}}}} > {\mu _{{\rm{LoS},k}}}\left( {r,\gamma } \right)U_{\rm{LoS},k}\left( {r,z_u} \right)} \right]} \right]} \\
    & + \sum\limits_{k = 1}^K {{\mathbb{E}_{r,U}}\left[ {P_{{\rm{NLoS}},k}^A\left( {r,z_u} \right)\mathbb{P}\left[ {{G_{{\rm{NLoS}}}} > {\mu _{{\rm{NLoS}}}}\left( {r,\gamma } \right)U_{\rm{NLoS},k}\left( {r,z_u} \right)} \right]} \right]}\\
    &\overset{(b)}{=} \sum\limits_{k = 1}^K \int_{{h_k}}^{ + \infty } {\mathbb{E}_U}\left[ {\mathbb{P}\left[ {{G_{{\rm{LoS}}}} > {\mu _{{\rm{LoS},k}}}\left( {r,\gamma } \right)U_{\rm{LoS},k}\left( {r,z_u} \right)} \right]} \right]  P_{{\rm{LoS}},k}^A\left( {r,z_u} \right)f_{{\rm{RLoS}},k}\left( {r,z_u} \right)\mathsf{d}r  \\
    & + \sum\limits_{k = 1}^K \int_{{h_k}}^{ + \infty } {\mathbb{E}_U}\left[ {\mathbb{P}\left[ {{G_{{\rm{NLoS}}}} > {\mu _{{\rm{NLoS}}}}\left( {r,\gamma } \right)U_{\rm{NLoS},k}\left( {r,z_u} \right)} \right]} \right]  P_{{\rm{NLoS}},k}^A\left( {r,z_u} \right)f_{{\rm{RNLoS}},k}\left( {r,z_u} \right)\mathsf{d}r ,
\end{split}
\end{equation}
\end{table*}
where $P_{{\rm{LoS}},k}^A\left( {r,z_u} \right)$, $P_{{\rm{NLoS}},k}^A\left( {r,z_u} \right)$ and $f_{{\rm{RQ}},k}\left( {r,z_u} \right)$ are given in (\ref{PALoSj}),  (\ref{PANLoSj}) and (\ref{sequation: PDF of tagged UAV}), respectively, $(a)$ is obtained by substituting $U_{Q,k}\left( {r,z_u} \right) = I_{Q,k}\left( {r,z_u} \right) + {\sigma ^2}$ and ${\mu_{Q,k}}\left(r,\gamma \right)$ are define in (\ref{miu}) into the former result, $(b)$ is obtained from the expectation of $r$. In order to get the final analytical result, the next steps are taken,
\begin{equation}
\begin{split}
\label{exact coverage probability_2}
    &{\mathbb{E}_U}\left[ { \mathbb{P}\left[ {{G_{{\rm{LoS}}}} > {\mu _{{\rm{LoS},k}}}\left( {r,\gamma } \right){U_{\rm{LoS},k}}\left( {r,z_u} \right)} \right]} \right] \\
    &\overset{(a)}{=}  {\mathbb{E}_U}\left[ {\frac{{{\Gamma _u}\left( {{m_{{\rm{LoS}}}},{m_{{\rm{LoS}}}}{\mu _{{\rm{LoS},k}}}\left( {r,\gamma } \right){U_{\rm{LoS},k}}\left( {r,z_u} \right)} \right)}}{{\Gamma \left( {{m_{{\rm{LoS}}}}} \right)}}} \right]\\
    &\overset{(b)}{=}  {\mathbb{E}_U}\Big[ \exp \left( { - {\mu _{{\rm{LoS},k}}}\left( {r,\gamma } \right)U\left( {r,z_u} \right)} \right)  \sum\limits_{n = 0}^{{m_{{\rm{LoS}}}} - 1} {\frac{{{{\left( {{\mu _{{\rm{LoS},k}}}\left( {r,\gamma } \right){U_{\rm{LoS},k}}\left( {r,z_u} \right)} \right)}^n}}}{{n!}}}  \Bigg]\\
    & = \sum\limits_{n = 0}^{{m_{{\rm{LoS}}}} - 1} {\frac{{{{\left( {{\mu _{{\rm{LoS},k}}}\left( {r,\gamma } \right)} \right)}^n}}}{{n!}}} {\mathbb{E}_U}\big[ \exp \left( { - {\mu _{{\rm{LoS},k}}}\left( {r,\gamma } \right){U_{\rm{LoS},k}}\left( {r,z_u} \right)} \right)
    {{\left( {{U_{\rm{LoS},k}}\left( {r,z_u} \right)} \right)}^n} \big]\\
    &\overset{(c)}{=}  \sum\limits_{n = 0}^{{m_{{\rm{LoS}}}} - 1} {{{\left[ {\frac{{{{\left( { - s} \right)}^n}}}{{n!}}\frac{{{\partial ^n}}}{{\partial {s^n}}}{\mathcal{L}_{U_{\rm{LoS},k}}}\left( {s,r,z_u} \right)} \right]}_{s = {\mu _{{\rm{LoS},k}}}\left( {r,\gamma } \right)}}},
\end{split}
\end{equation}
where $(a)$ follows the \ac{CCDF} of the Gamma distribution ${\overline F _G}\left( g \right) = \frac{{{\Gamma _u}\left( {m,mg} \right)}}{{\Gamma \left( m \right)}}$, where ${\Gamma _u}\left( {m,mg} \right) = \int_{mg}^{ + \infty } {{t^{m - 1}}{e^{ - t}}dt} $ is the upper incomplete Gamma function, and $(b)$ follows the definition $\frac{{{\Gamma _u}\left( {m,mg} \right)}}{{\Gamma \left( m \right)}} = \exp \left( { - g} \right)\sum\limits_{n = 0}^{m - 1} {\frac{{{g^n}}}{{n!}}} $ \cite{LT_Nakagami}, by the linearity of the expectation operator and
\begin{equation}
\begin{split}
   & {\mathbb{E}_U}\left[ {\exp \left( { - s{U_{\rm{LoS},k}}\left( {r,z_u} \right)} \right){U_{\rm{LoS},k}}{{\left( {r,z_u} \right)}^n}} \right] = {\left( { - 1} \right)^n}\frac{{{\partial ^n}}}{{\partial {s^n}}}{\mathcal{L}_{U_{\rm{LoS},k}}}\left( {s,r,z_u} \right),  
\end{split}
\end{equation}
$(c)$ is obtained. The steps of NLoS UAVs are similar to that of LoS UAVs, therefore omitted here.

\section{Proof of Theorem~\ref{approximate  theorem}}\label{app:approximate theorem}
Because the first several steps of the proof of approximate coverage probability are similar to that of exact coverage probability, we start from formulation (\ref{exact coverage probability_2}) step $(a)$,
\begin{equation}
\begin{split}
    &{\mathbb{E}_{U}}\left[ {\frac{{{\Gamma _u}\left( {{m_{{\rm{LoS}}}},{m_{{\rm{LoS}}}}{\mu _{{\rm{LoS},k}}}\left( {r,\gamma } \right){U_{\rm{LoS},k}}\left( {r\left| {{r_u}} \right.} \right)} \right)}}{{\Gamma \left( {{m_{{\rm{LoS}}}}} \right)}}} \right] \\
    &\overset{(a)}{=} 1 - {\mathbb{E}_U}\left[ {\frac{{{\Gamma _l}\left( {{m_{{\rm{LoS}}}},{m_{{\rm{LoS}}}}{\mu _{{\rm{LoS},k}}}\left( {r,\gamma } \right){U_{\rm{LoS},k}}\left( {r\left| {{r_u}} \right.} \right)} \right)}}{{\Gamma \left( {{m_{{\rm{LoS}}}}} \right)}}} \right]\\
    &\overset{(b)}{\approx} 1 - {\mathbb{E}_U}\left[ {{{\left( {1 - \exp \left( { - {\beta _{{\rm{LoS}}}}{\mu _{{\rm{LoS},k}}}\left( {r,\gamma } \right){U_{\rm{LoS},k}}\left( {r,z_u} \right)} \right)} \right)}^{{m_{{\rm{LoS}}}}}}} \right]\\
    &\overset{(c)}{=} {\mathbb{E}_U} \Bigg[ \sum\limits_{n = 1}^{{m_{{\rm{LoS}}}}} {\binom{m_{\rm{LoS}}}{n}} {{\left( { - 1} \right)}^{n + 1}} \exp \left( { - n{\omega _{{\rm{LoS}}}}{\mu _{{\rm{LoS},k}}}\left( {r,\gamma } \right){U_{\rm{LoS},k}}\left( {r,z_u} \right)} \right) \Bigg]\\
    & = \sum\limits_{n = 1}^{{m_{{\rm{LoS}}}}} {\binom{m_{\rm{LoS}}}{n}{{\left( { - 1} \right)}^{n + 1}}{\mathcal{L}_{U_{\rm{LoS},k}}}\left( {n{\omega _{{\rm{LoS}}}}{\mu _{{\rm{LoS},k}}}\left( {r,\gamma } \right)} \right)},
\end{split}
\end{equation}
where ${\mathcal{L}_{U_{\rm{LoS},k}}}\left( s,r,k\right)$ is given in (\ref{L_U}), and $s = n{\omega _{{\rm{LoS}}}}{\mu _{{\rm{LoS},k}}}\left( {r,\gamma } \right)$,  $\Gamma_{l}\left(m,mg\right) = \int_0^{mg} {{t^{m - 1}}{e^{ - t}}dt}$ in step $(a)$ is the lower incomplete Gamma function, which satisfies $\frac{{{\Gamma _u}\left( {m,mg} \right)}}{{\Gamma \left( m \right)}} = 1 - \frac{{{\Gamma _l}\left( {m,mg} \right)}}{{\Gamma \left( m \right)}}$. $(b)$ follows from the tight approximation to coverage probability, where ${\omega _{{\rm{LoS}}}} = {\left( {{m_{{\rm{LoS}}}}!} \right)^{\frac{{ - 1}}{{{m_{{\rm{LoS}}}}}}}}$. It has been proved in \cite{bound_of_Gamma} that the tighter upper bound provides an accurate approximation of the CDF of the Gamma distribution, which is bounded by
\begin{equation}
\label{upper bound}
    {\left( {1 - {e^{ - {\omega _1}mg}}} \right)^m} < \frac{{{\Gamma _l}\left( {m,mg} \right)}}{{\Gamma \left( m \right)}} < {\left( {1 - {e^{ - {\omega _2}mg}}} \right)^m},
\end{equation}
where $m \ne 1$, and
\begin{equation}
{\omega _1} = \left\{ {\begin{array}{*{20}{c}}
{1,}&{{\rm{if\,\,}}m > 1}\\
{{{\left( {m!} \right)}^{\frac{{ - 1}}{m}}},}&{{\rm{if\,\,}}m < 1}
\end{array}} \right.{\omega _2} = \left\{ {\begin{array}{*{20}{c}}
{{{\left( {m!} \right)}^{\frac{{ - 1}}{m}}},}&{{\rm{if\,\,}}m > 1}\\
{1,}&{{\rm{if\,\,}}m < 1}
\end{array}} \right.,
\end{equation}
and step $(c)$ is given by the binomial theorem, and it is necessary to assume that ${m_{{\rm{LoS}}}}$ is an integer.

\bibliographystyle{IEEEtran} 
\bibliography{references}

\end{document}